\documentclass[%
 reprint,
superscriptaddress,
 amsmath,amssymb,
 aps,
prb,
]{revtex4-1}

\usepackage{graphicx}
\usepackage{dcolumn}
\usepackage{bm}
\usepackage{hyperref}

\usepackage{mhchem}
\usepackage{subcaption}
\usepackage[export]{adjustbox}

\begin{document}

\title{Radiation tolerance of two-dimensional material-based devices for space applications}

\author{Tobias Vogl}
\email{To whom correspondence should be addressed:\\ Ping Koy Lam (ping.lam@anu.edu.au), or Tobias Vogl\\ (tobias.vogl@anu.edu.au)}
\affiliation{Centre for Quantum Computation and Communication Technology, Department of Quantum Science, Research School of Physics and Engineering, The Australian National University, Acton ACT 2601, Australia}
\author{Kabilan Sripathy}
\affiliation{Centre for Quantum Computation and Communication Technology, Department of Quantum Science, Research School of Physics and Engineering, The Australian National University, Acton ACT 2601, Australia}
\author{Ankur Sharma}
\affiliation{Research School of Engineering, The Australian National University, Acton ACT 2601, Australia}
\author{Prithvi Reddy}
\affiliation{Laser Physics Centre, Research School of Physics and Engineering, The Australian National University, Acton, ACT 2601, Australia}
\author{James Sullivan}
\affiliation{Plasma Research Laboratory, Research School of Physics and Engineering, The Australian National University, Canberra ACT 2601, Australia}
\author{Joshua R. Machacek}
\affiliation{Plasma Research Laboratory, Research School of Physics and Engineering, The Australian National University, Canberra ACT 2601, Australia}
\author{Linglong Zhang}
\affiliation{Research School of Engineering, The Australian National University, Acton ACT 2601, Australia}
\author{Fouad Karouta}
\affiliation{Australian National Fabrication Facility, Research School of Physics and Engineering, The Australian National University, Acton, ACT 2601, Australia}
\author{Ben C. Buchler}
\affiliation{Centre for Quantum Computation and Communication Technology, Department of Quantum Science, Research School of Physics and Engineering, The Australian National University, Acton ACT 2601, Australia}
\author{Marcus W. Doherty}
\affiliation{Laser Physics Centre, Research School of Physics and Engineering, The Australian National University, Acton, ACT 2601, Australia}
\author{Yuerui Lu}
\affiliation{Research School of Engineering, The Australian National University, Acton ACT 2601, Australia}
\author{Ping Koy Lam}
\email{To whom correspondence should be addressed:\\ Ping Koy Lam (ping.lam@anu.edu.au), or Tobias Vogl\\ (tobias.vogl@anu.edu.au)}
\affiliation{Centre for Quantum Computation and Communication Technology, Department of Quantum Science, Research School of Physics and Engineering, The Australian National University, Acton ACT 2601, Australia}

\date{\today}

\begin{abstract}
Characteristic for devices based on two-dimensional materials are their low size, weight and power requirements. This makes them advantageous for use in space instrumentation, including photovoltaics, batteries, electronics, sensors and light sources for long-distance quantum communication. Here, we present for the first time a comprehensive study on combined radiation effects in earth's atmosphere on various devices based on these nanomaterials. Using theoretical modeling packages, we estimate relevant radiation levels and then expose field-effect transistors, single-photon sources and monolayers as building blocks for future electronics to gamma-rays, protons and electrons. The devices show negligible change in performance after the irradiation, suggesting robust suitability for space use. Under excessive $\gamma$-radiation, however, monolayer WS$_2$ showed decreased defect densities, identified by an increase in photoluminescence, carrier lifetime and a change in doping ratio proportional to the photon flux. The underlying mechanism was traced back to radiation-induced defect healing, wherein dissociated oxygen passivates sulfur vacancies.
\end{abstract}

\maketitle


In the near future, quantum tunneling will set a hard limit to further miniaturization of silicon-based electronics. Research on alternative materials, however, demonstrated fabrication beyond this limit\cite{10.1038/nnano.2011.56,10.1038/nnano.2012.21}. Of particular interest are monolayered two-dimensional (2D) materials such as Graphene\cite{10.1126/science.1102896} and transition metal dichalcogenides (TMDs) like MoS$_2$\cite{PhysRevLett.105.136805}. Record electron mobility in 2D materials has enabled multiple technology demonstrations of atomically thin transistors\cite{10.1038/nnano.2010.89,10.1038/nnano.2010.279,doi:10.1021/nn501723y,10.1038/nnano.2016.115,10.1126/science.aah4698}. Furthermore, due to their semiconducting bandstructure, TMDs have applications in optoelectronics and photonics\cite{10.1088/1674-4926/38/3/031002}. Their intrinsically low size, weight and power (SWaP) requirements and their chemical stability make 2D material-based devices a promising candidate for space instrumentation. Beyond integrated electronics, 2D materials in space technology can be utilized for solar cells\cite{10.1038/s41699-018-0049-3}, batteries\cite{10.1039/C6TA09831B}, sensors as well as non-classical light sources for long-distance quantum communication\cite{nnano.2015.242}. The quantum emission from point defects in 2D materials have desirable properties for single-photon sources, as they can be easily integrated with photonic networks, have an intrinsic out-coupling efficiency of unity and offer long-term stable, high luminosity single-photons at room temperature\cite{nnano.2015.242,10.1021/acsphotonics.8b00127,0022-3727-50-29-295101,doi:10.1021/acsnano.6b03602,doi:10.1021/acsami.6b09875}. An ideal single-photon source can enhance the data communication rates of satellite-based quantum key distribution\cite{10.1038/nature23655}.\\
\indent While 2D materials offer great opportunities for space missions, their current low technological readiness level (TRL) restricts deployment (current state-of-the-art is TRL 3-4). In addition to further device development, 2D materials need to be certified for the harsh conditions of space. Space qualification studies usually consist of vacuum and thermal cycling, vibration and shock tests as well as exposure to radiation\cite{GSFC-STD-7000}. Vibration or shock will not pose a threat for nanomaterials and vacuum and thermal cycling is routinely done in experiments\cite{PhysRevB.93.205423,doi:10.1021/acsphotonics.7b00086,PhysRevB.98.081414}. Of particular interest, however, is the effect of radiation on 2D materials. While radiation effects on the electrical properties of Graphene have been studied\cite{10.1088/1674-1056/20/8/086102,doi:10.1063/1.4963782}, less is known about these effects on TMDs and other 2D materials\cite{10.1002/pssa.201670681}. In particular, no study investigates the effect of radiation on optical characteristics of 2D materials. Moreover, there exists no comprehensive study on the effects of combined radiation types on properties of various devices in the context of space certification. The damage caused by high-energy particles and photons in the X- and $\gamma$-ray range is of major concern for all spacecraft, especially as weight restrictions limit shielding options. While testing directly in a space environment as planned for Graphene is possible\cite{Milliron}, a more practical way is to replicate space environments on earth.\\
\indent Here, we present a comprehensive study on the effects of radiation in the atmosphere on various devices based on 2D materials: Single-photon sources based on defects in hexagonal boron nitride (hBN) and field-effect transistors (FETs) based on monolayer MoS$_2$ and WSe$_2$. We also tested MX$_2$ monolayers (M = Mo, W and X = S, Se) in their native state as basic building blocks for future electronics and optoelectronics. All devices are influenced by their electrical and/or optical properties (quantum emitters in hBN are dependent on the piezoelectric environment of hBN\cite{10.1364/OPTICA.5.001128}). It is possible that low-energy radiation could change the charge state of the defects in hBN, causing them to enter a dark state. Conversely, high-energy radiation could create new defects in the crystal lattice, which, if close to the quantum emitter could make the emitter unusable. With respect to the FETs, the radiation could change the carrier density, which alters their performance.\\
\indent We start with modeling radiation levels in the thermosphere using the SPace ENVironment Information System (SPENVIS), software provided by the European Space Agency\cite{spenvis}. With the knowledge from the simulations, we expose our devices to the most common radiation types in the low earth orbit (LEO): Gamma-rays as well as energetic protons and electrons. We look at isolated effects (effects solely caused by a specific type of radiation) and combined effects by exposing devices to all three types of radiation. For each test we fully characterize all devices back-to-back, shortly before and after the exposure. Any changes caused by the radiation are studied and the underlying mechanisms were then traced back. Furthermore, the dynamics of the interaction between matter and radiation are modeled using Monte Carlo simulations and density functional theory calculations.

\section*{Results}
\subsubsection*{Radiation levels in LEO}
\begin{figure*}[t!]
  \centering
  \includegraphics[width=0.49\linewidth,keepaspectratio,valign=t]{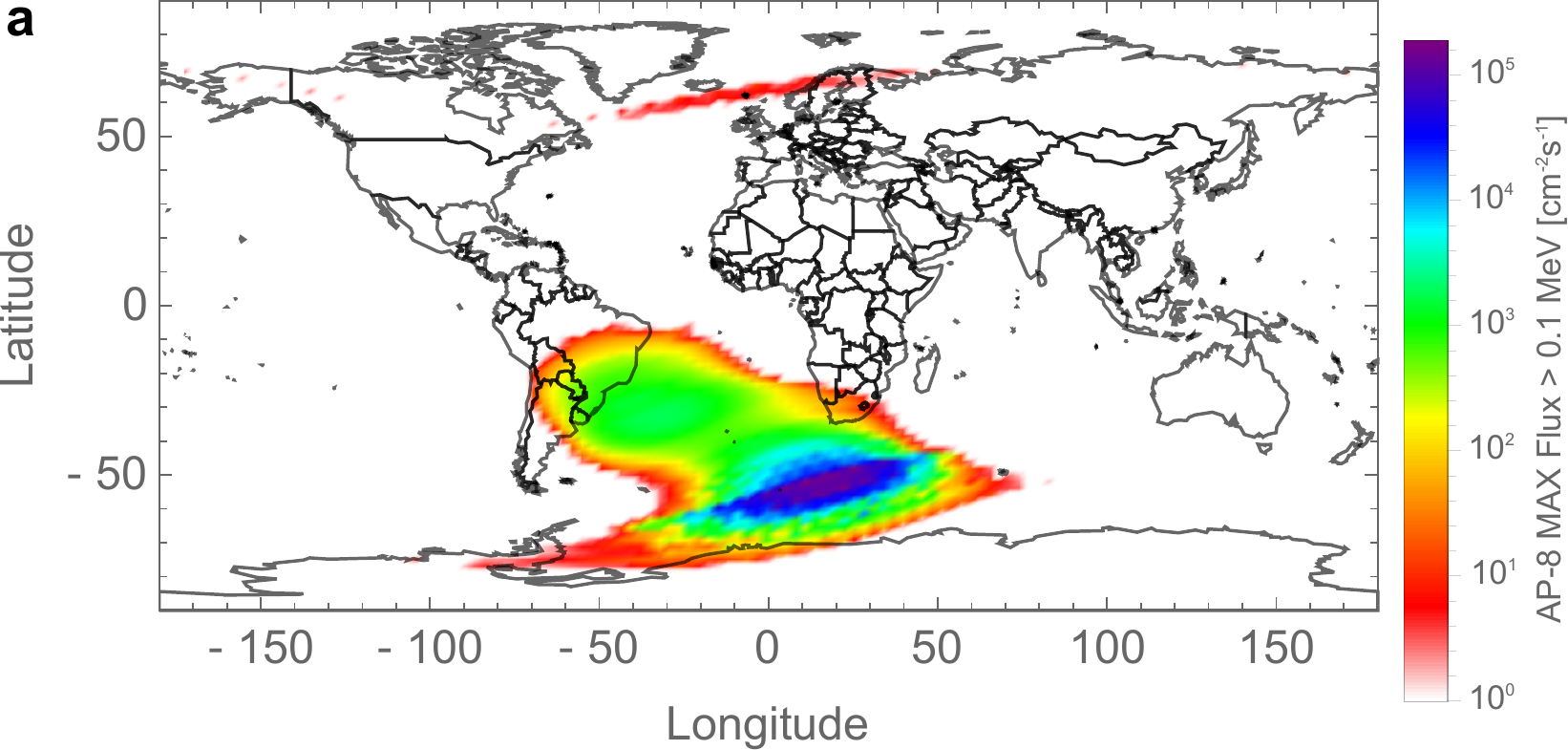}
  \includegraphics[width=0.49\linewidth,keepaspectratio,valign=t]{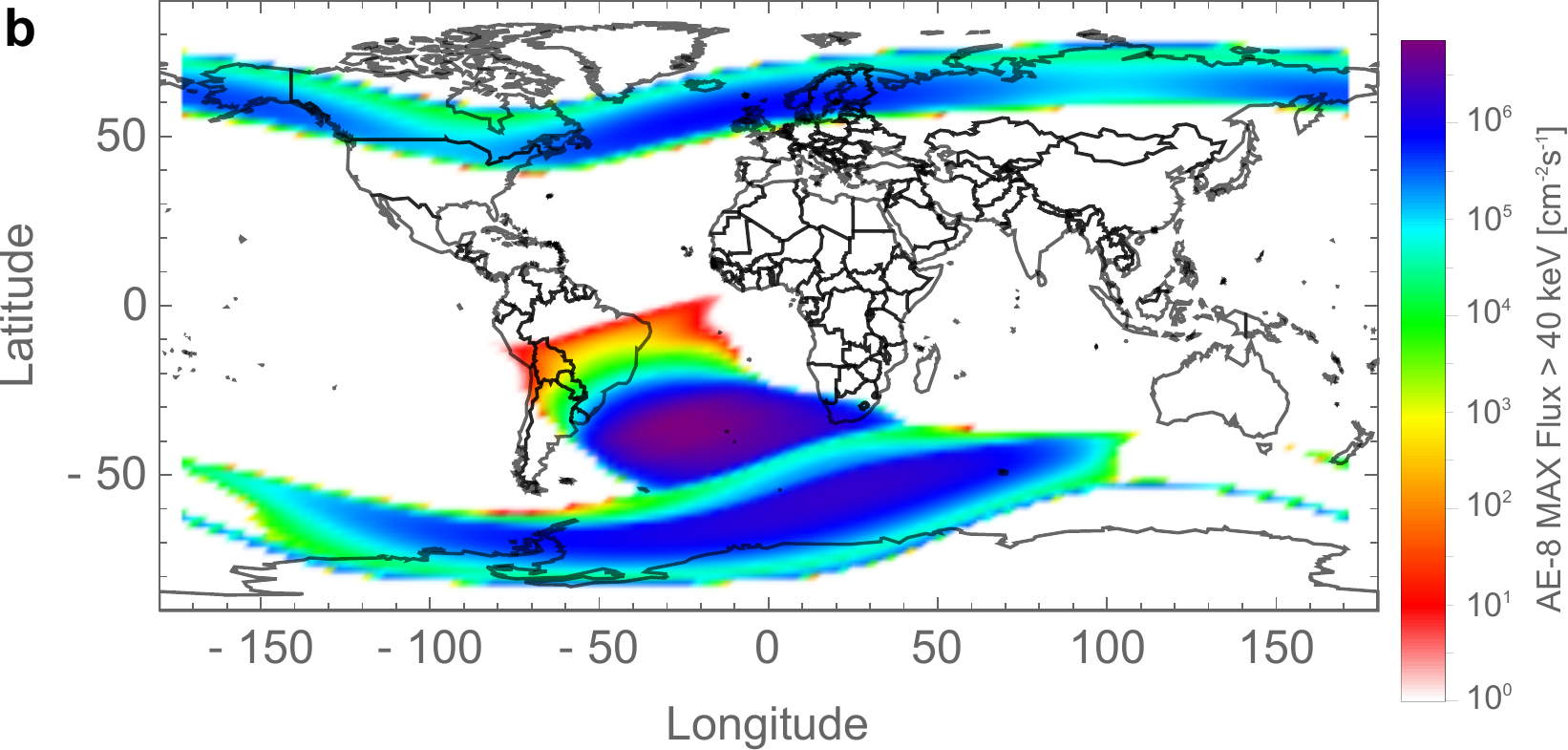}\\
  \vspace{0.1cm}
  \includegraphics[width=0.49\linewidth,keepaspectratio,valign=t]{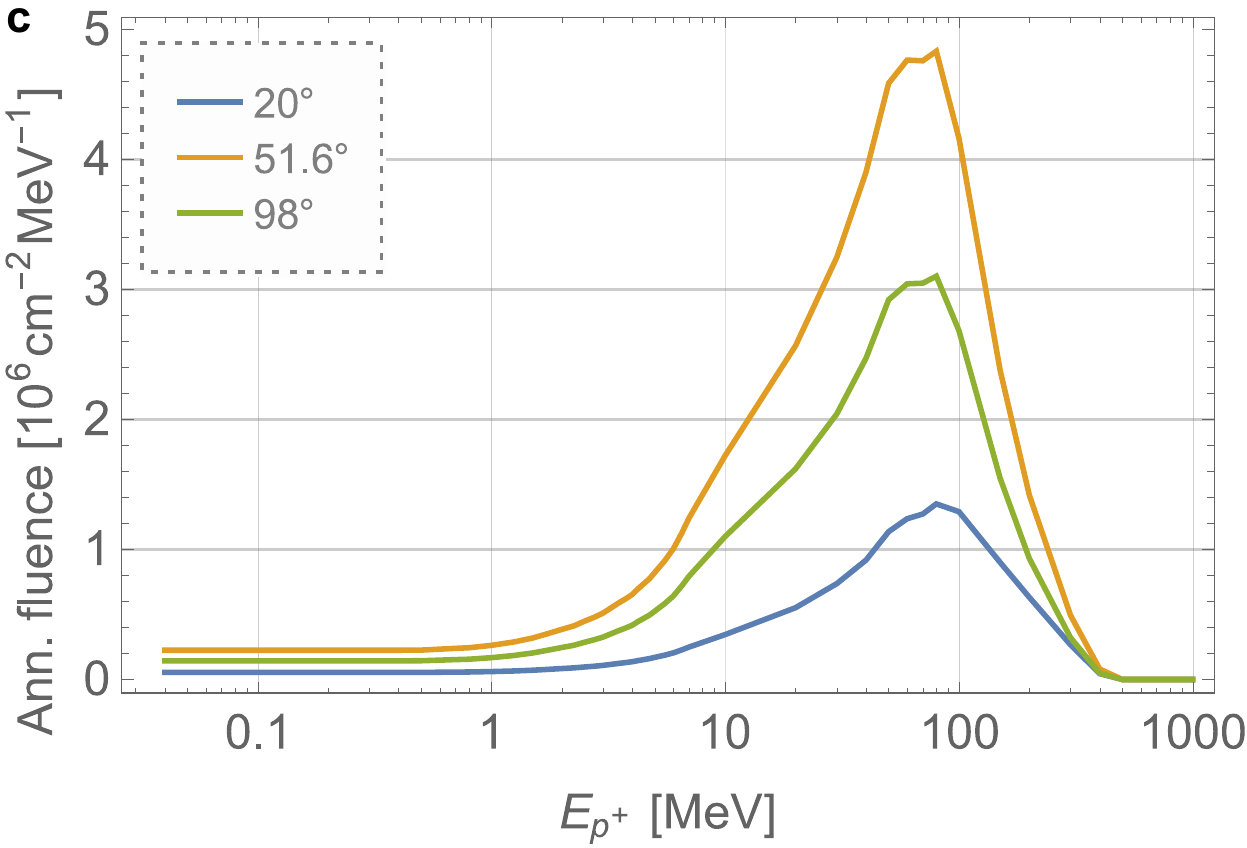}
  \includegraphics[width=0.49\linewidth,keepaspectratio,valign=t]{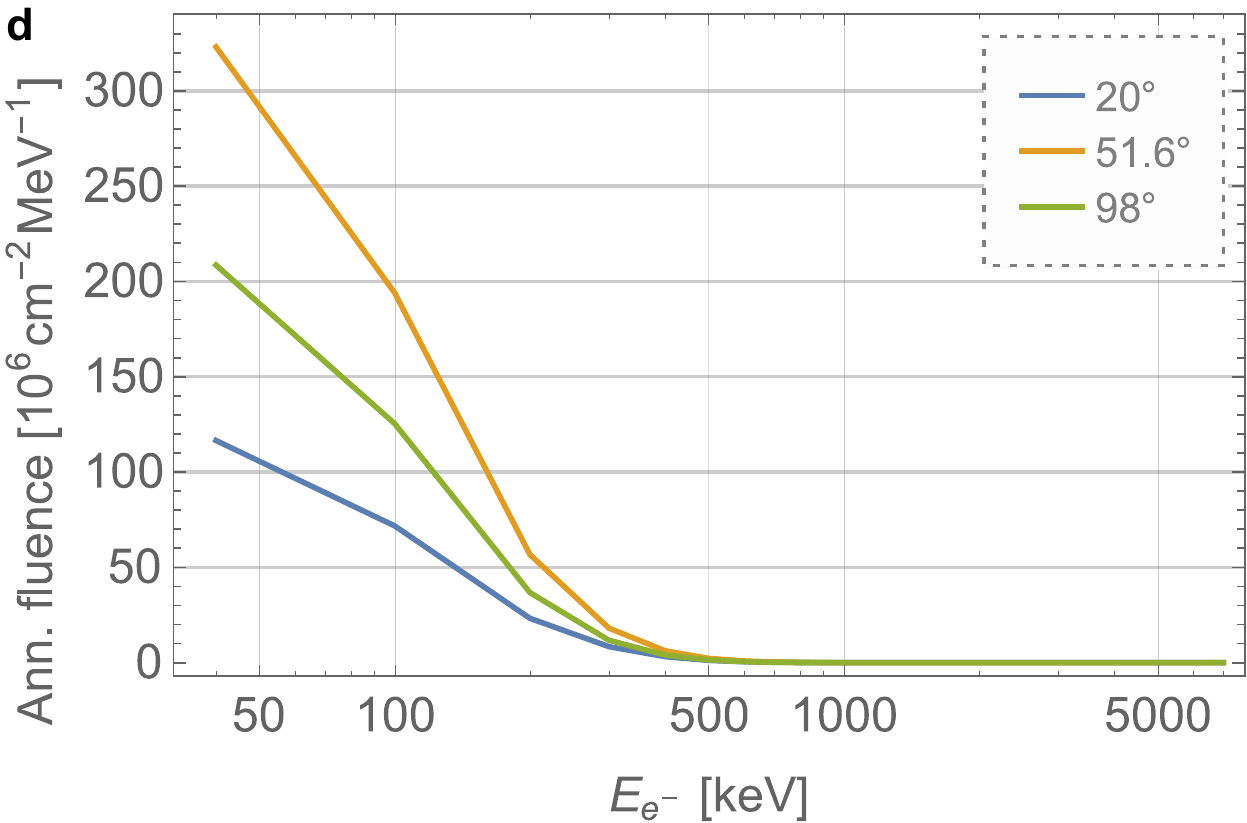}
\caption{\textbf{Space environment.} Geographical distribution of the trapped \textbf{(a)} proton and \textbf{(b)} electron flux at 500$\,$km altitude. Integrated annual \textbf{(c)} proton and \textbf{(d)} electron fluence after $1.85\,$mm of Al shielding for typical orbital inclinations.}
\label{fig:1}
\end{figure*}

The earth is protected from solar wind and cosmic particles by its magnetic field. As a result, high-energy protons and electrons are trapped on trajectories oscillating between both magnetic poles in the so-called Van Allen belts. While essential for life on earth, the trapped particle belts pose great threat to any spacecraft orbiting through these trapped particle belts. Near the magnetic poles the inner belt can extend down to altitudes of 200$\,$km. Due to misalignment of the magnetic dipole and rotation axis of the earth, this appears as the South Atlantic Anomaly (SAA, see Figure $\ref{fig:1}$(a,b)). Because of this inhomogeneity, the total radiation dosage is strongly dependent on the orbital inclination. Thus, we calculated the particle spectra for different spacecraft trajectories with inclinations of $20^\circ$ (in the following defined as equatorial orbit), $51.6^\circ$ (orbit of the International Space Station, ISS) and $98^\circ$ (in the following defined as polar orbit) for $500\,$km altitude and average over the full orbit. In general, the energy spectrum for protons in LEO ranges from $100\,$keV to $400\,$MeV, while for electrons it ranges from $40\,$keV to $7\,$MeV. Low-energy particles are typically absorbed by the walls of the spacecraft, which acts as a non-ideal high-pass filter. High-energy ions, however, loose energy during their interaction with the shielding material and thus, the lower ends of the spectra are always finite unless every charge carrier is stopped (e.g. for large shielding thicknesses). The shielded flux spectra for protons and electrons after $1.85\,$mm of Al shielding and integrated over a one year mission is shown in Figure $\ref{fig:1}$(c,d). Surprisingly, the polar orbit does not have the highest fluence, as spacecrafts with $51.6^\circ$ inclination spend more time in the SAA than spacecrafts with $98^\circ$ inclination, similarly for protons and electrons. A spectral distribution with an absence of low-energy protons, as shown in Figure $\ref{fig:1}$(c), is advantageous, because only low-energy particles can deposit significant amounts of energy into the payload. It should be mentioned, that the electrons in Figure $\ref{fig:1}$(d) do not originate from trapped electrons in the Van Allen belt, but rather are secondary electrons produced via ionizing interactions of high-energy protons with the Al atoms in the shielding material.\\
\indent While the particle fluence spectra are directly accessible through SPENVIS, similar tools for $\gamma$-rays do not exist. Gamma-rays will most likely originate either directly from the sun or from radioactive decay of trapped particles in the radiation belts. For our study we use data from the CORONAS-I satellite\cite{bucik2000,bucik1999}, which mapped the $\gamma$-ray flux above the earth at 500$\,$km altitude. A summary of reported flux values is given in Table $\ref{table1}$.

\subsubsection*{Device fabrication and characterization}
\begin{figure*}[t!]
\centering
  \includegraphics[width=0.32\linewidth,keepaspectratio,valign=t]{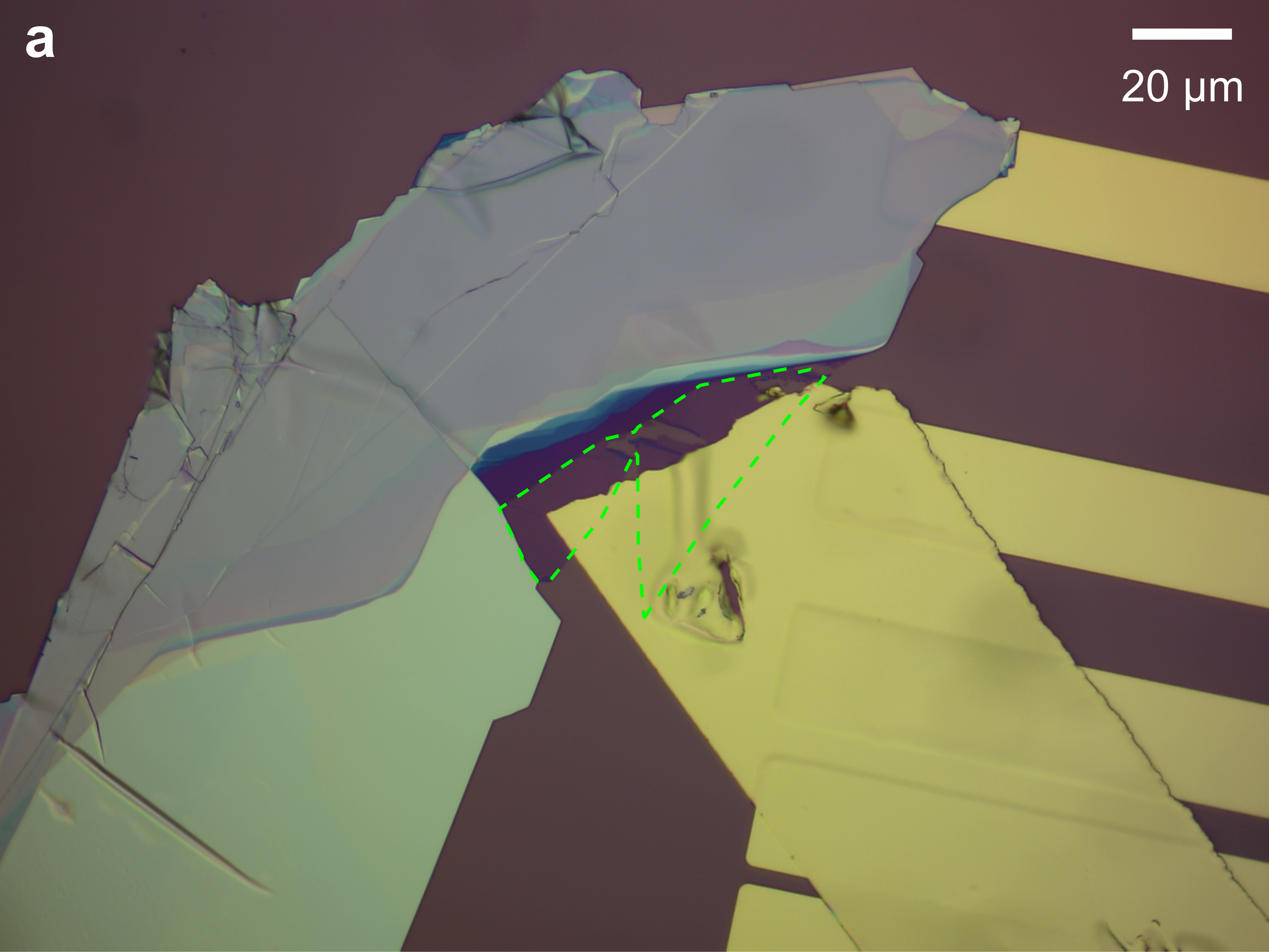}
  \includegraphics[width=0.32\linewidth,keepaspectratio,valign=t]{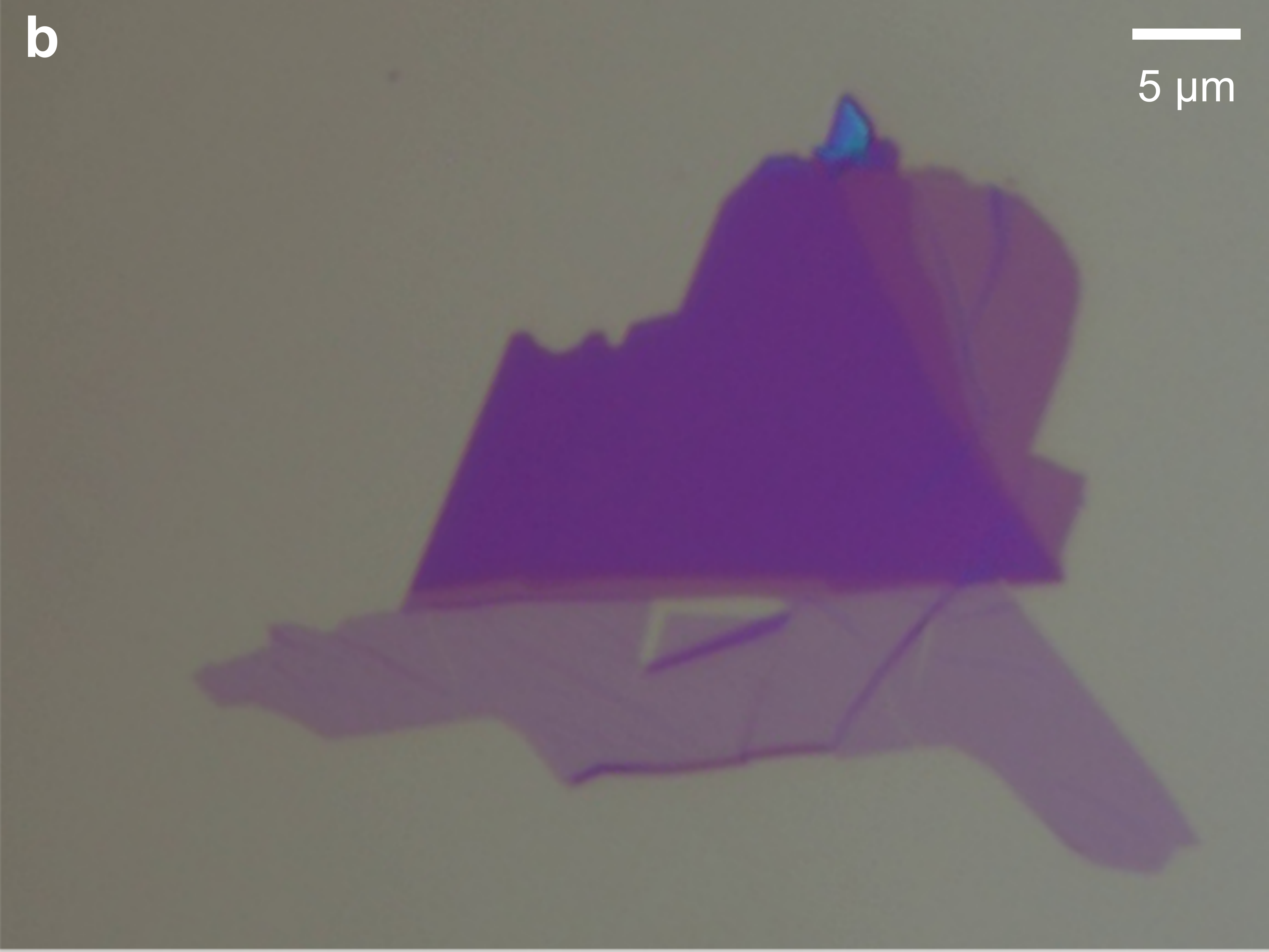}
  \includegraphics[width=0.32\linewidth,keepaspectratio,valign=t]{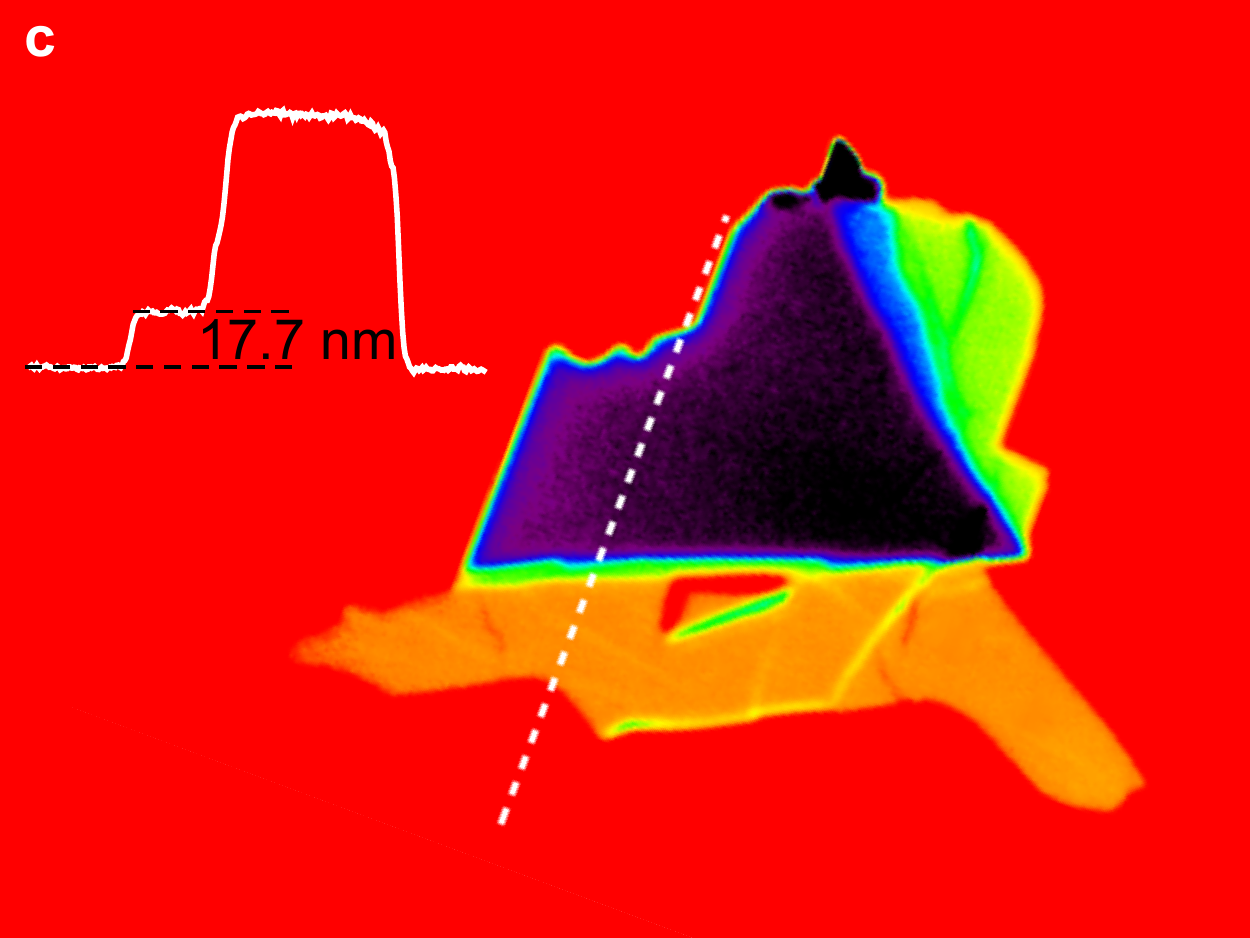}
\caption{\textbf{Fabrication. (a)} Microscope image of a MoS$_2$ FET device under 500$\times$ magnification. The monolayer is framed within the green dashed line. \textbf{(b,c)} Microscope and PSI images of the WS$_2$ monolayer presented in the main text. The inset in \textbf{(c)} shows the OPD along the white dashed line. The monolayer has an OPD of 17.7$\,$nm, which corresponds to a physical thickness of 0.66$\,$nm.}
\label{fig:2}
\end{figure*}

All 2D crystals have been mechanically exfoliated from bulk crystal to a viscoelastic gel foil. Monolayer TMDs and multilayer hBN have been identified by contrast-enhanced microscopy and transferred by dry contact to Si chips capped with a layer of $262\,$nm thermally grown SiO$_2$ (Si/SiO$_2$ substrate). Unless stated otherwise, all optical and electrical measurements have been carried out at room temperature (RT). While more than 80 devices were investigated throughout this study, herein we only show exemplary results and average over the full data set. More data is shown in the Supplementary Information.\\
\indent The hBN flakes have been treated with an oxygen plasma and successively rapidly thermally annealed\cite{10.1021/acsphotonics.8b00127}. The oxygen plasma creates point defects in the crystal lattice, which act as trapping sites for localized excitons. The single-photon emitters formed in this way were located and characterized using a confocal microphotoluminescence ($\mu$PL) system equipped with an ultrashort-pulsed laser for time-resolved measurements (see Methods). The emitters are excited off-resonantly at 522$\,$nm, less than half of the band gap of hBN ($E_g=6\,$eV\cite{10.1038/nphoton2015.77}), preventing two-photon absorption. For the quantum emitters we measured the spectrum, excited state lifetime and second-order correlation function. The latter was measured with a Hanbury Brown and Twiss (HBT)-type interferometer (see Methods).\\
\indent The atomically-thin FET devices (see Figure $\ref{fig:2}$(a)) were fabricated using pre-patterned gold electrode substrates and mechanical transfer of the monolayer onto the electrodes (see Methods). The device ON/OFF performance characterization was done using the standard back gate sweep from $-60$ to $+60\,$V at different biases between source and drain. Also, the conventional performance $I$-$V$ curves of the device were recorded at various back gate voltages in the ON regime of the functional FET device.\\
\indent Since monolayered 2D materials have often been proposed as candidates for the post-silicon age, we also tested blank monolayered TMDs (MX$_2$). After transfer to the Si/SiO$_2$ substrate (see Figure $\ref{fig:2}$(b)), the monolayer thickness is confirmed by phase-shift interferometry (PSI), with the corresponding PSI image shown in Figure $\ref{fig:2}$(c). In this case, the WS$_2$ crystal has an optical path length difference (OPD) of 17.7$\,$nm. With rigorous coupled-wave analysis (RCWA) simulations\cite{10.1038/lsa.2016.46}, the OPD can be converted to a physical thickness of 0.66$\,$nm, matching well atomic force microscope (AFM) measurements\cite{doi:10.1021/nl3026357}. We characterized each flake optically with the $\mu$PL setup in terms of emission spectrum (averaged over the full monolayer), carrier lifetime and power saturation. The carrier lifetime data, deconvoluted from the system response, is fitted with a bi-exponential from which radiative and non-radiative decay time $\tau_\text{r}$ and $\tau_\text{nr}$ can be extracted. Every flake is scanned with a $1\,\mu$m grid and a spectrum is recorded at each point. In order to gather enough statistics, a total of 49 monolayer flakes with areas ranging from 60 to $1290\,\mu$m$^2$ have been characterized.

\subsubsection*{Gamma-ray tests}
\begin{figure*}[t!]
\centering
  \includegraphics[width=0.36\linewidth,keepaspectratio,valign=t]{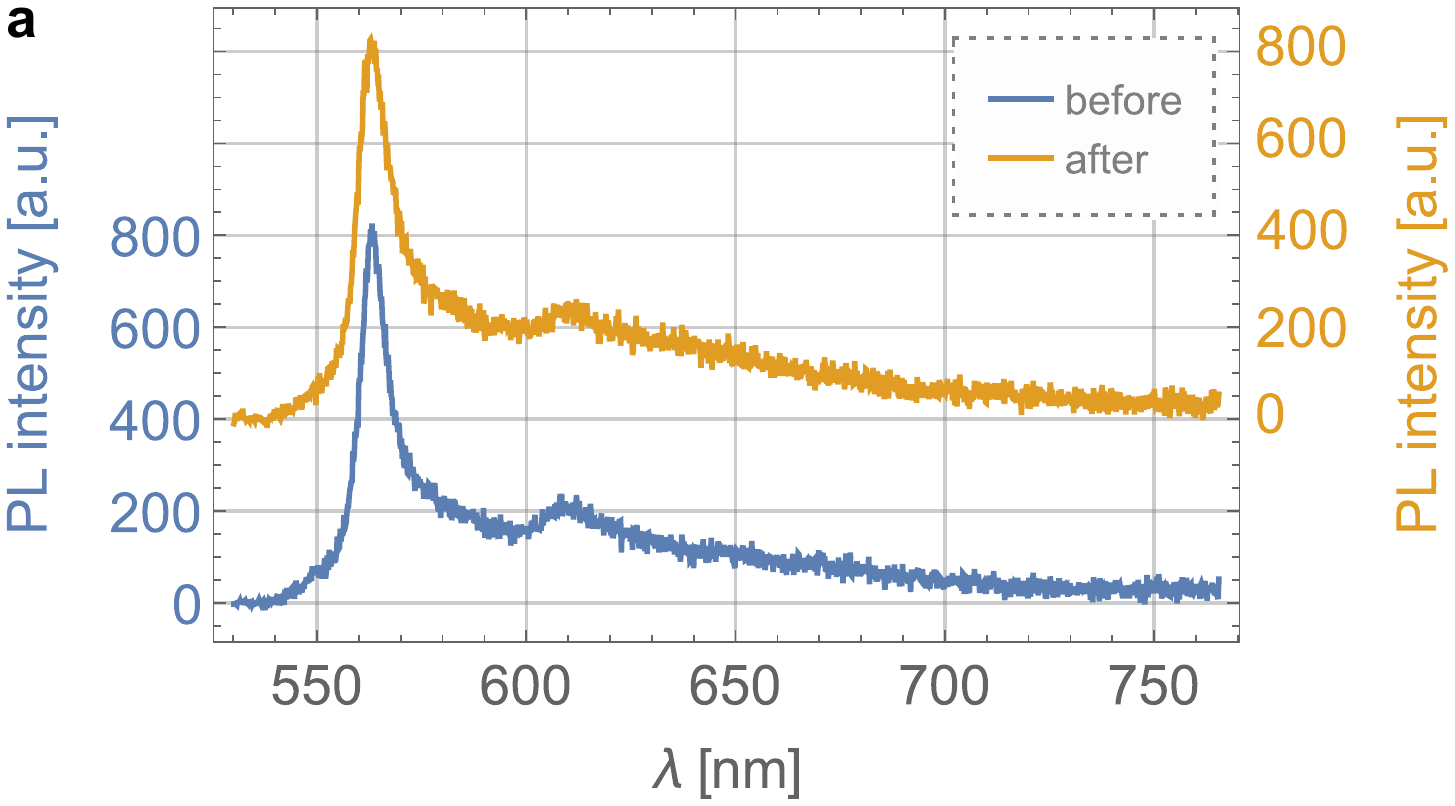}
  \includegraphics[width=0.3\linewidth,keepaspectratio,valign=t]{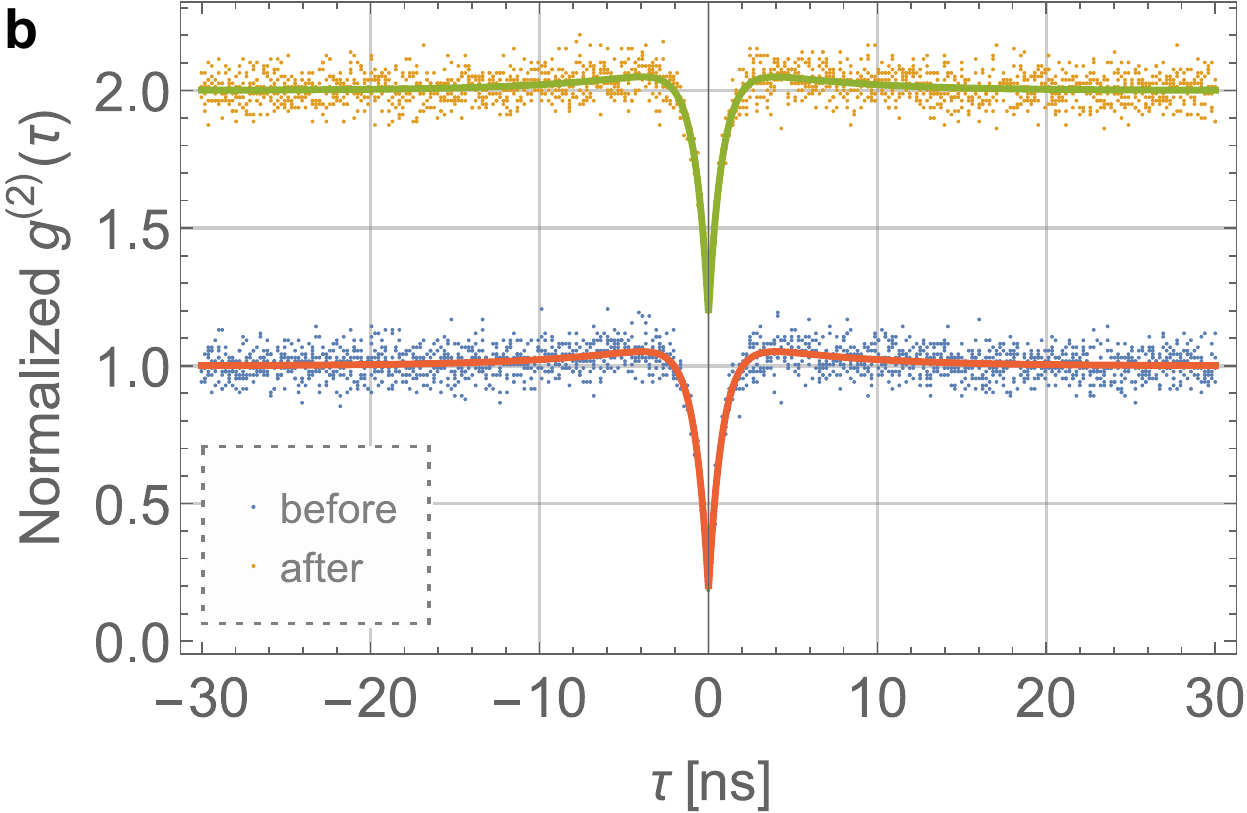}
  \includegraphics[width=0.3\linewidth,keepaspectratio,valign=t]{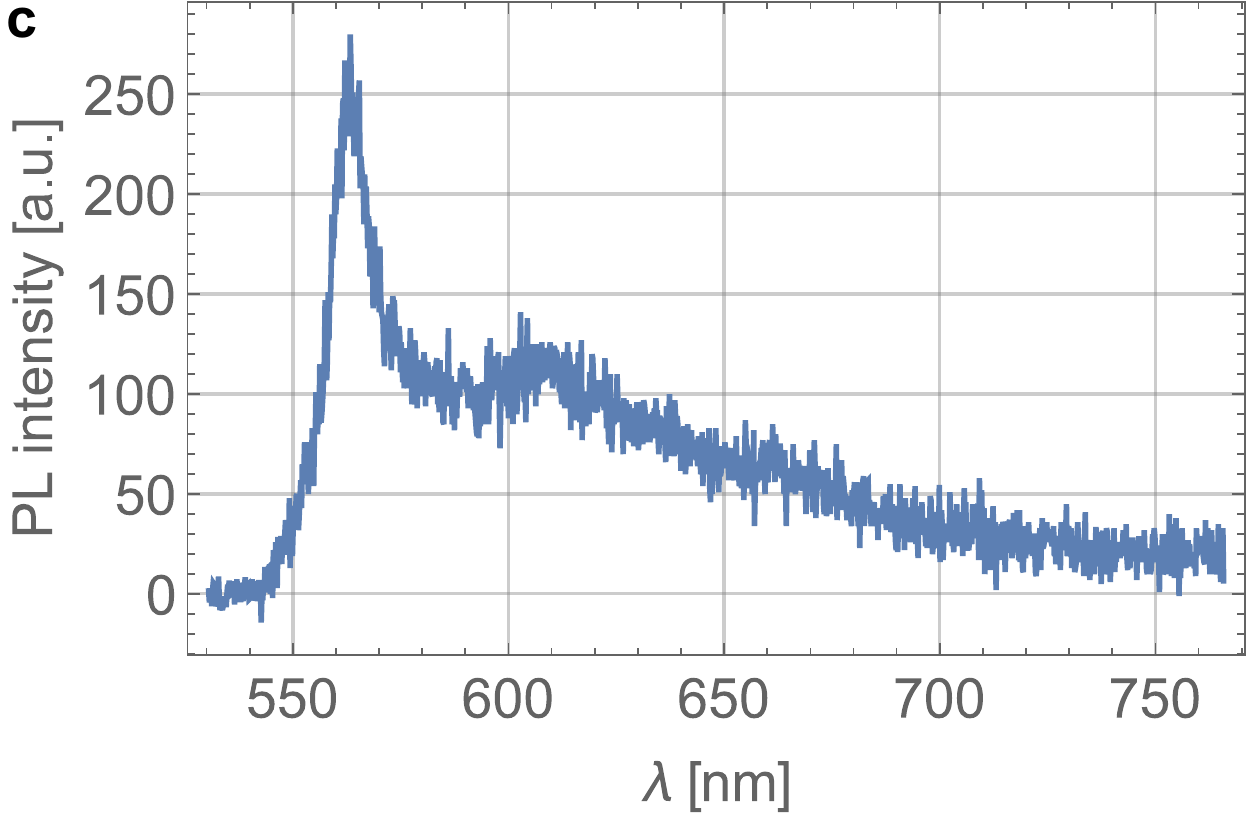}\\
  \vspace{0.1cm}
  \includegraphics[width=0.32\linewidth,keepaspectratio,valign=t]{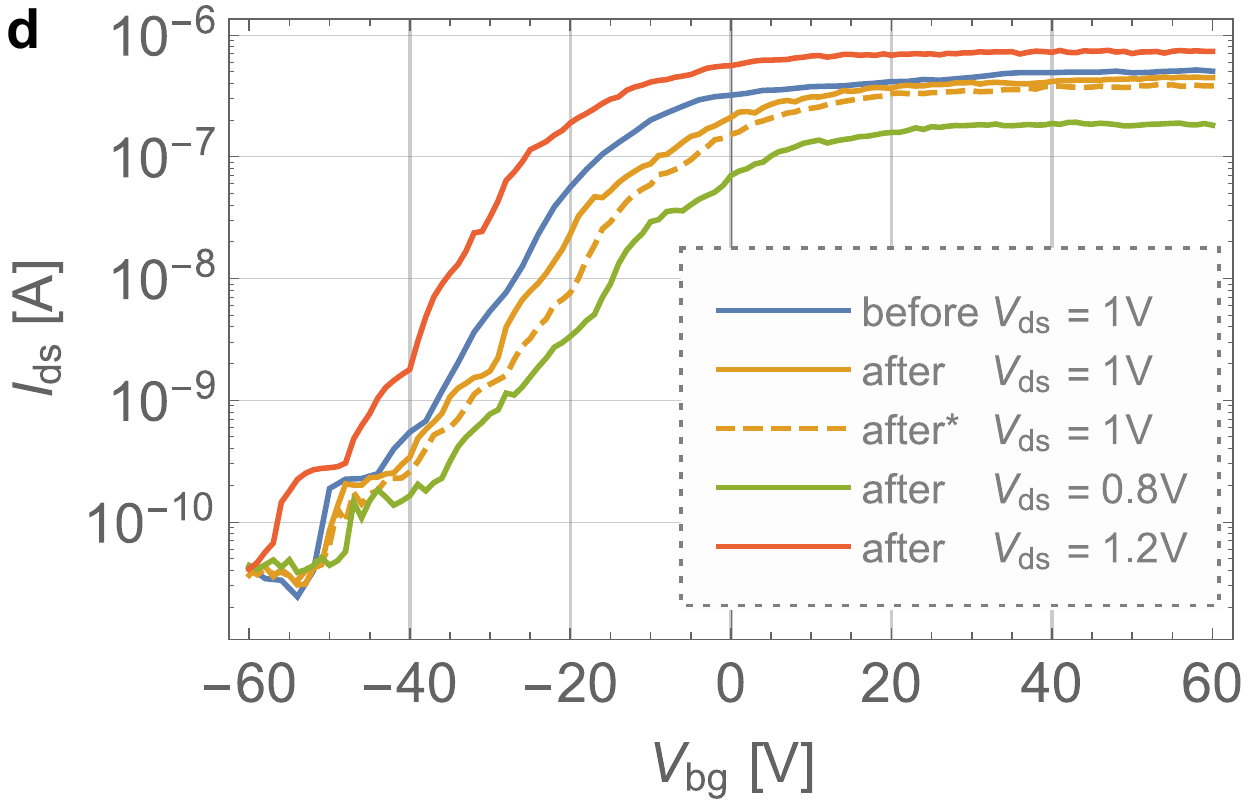}
  \includegraphics[width=0.32\linewidth,keepaspectratio,valign=t]{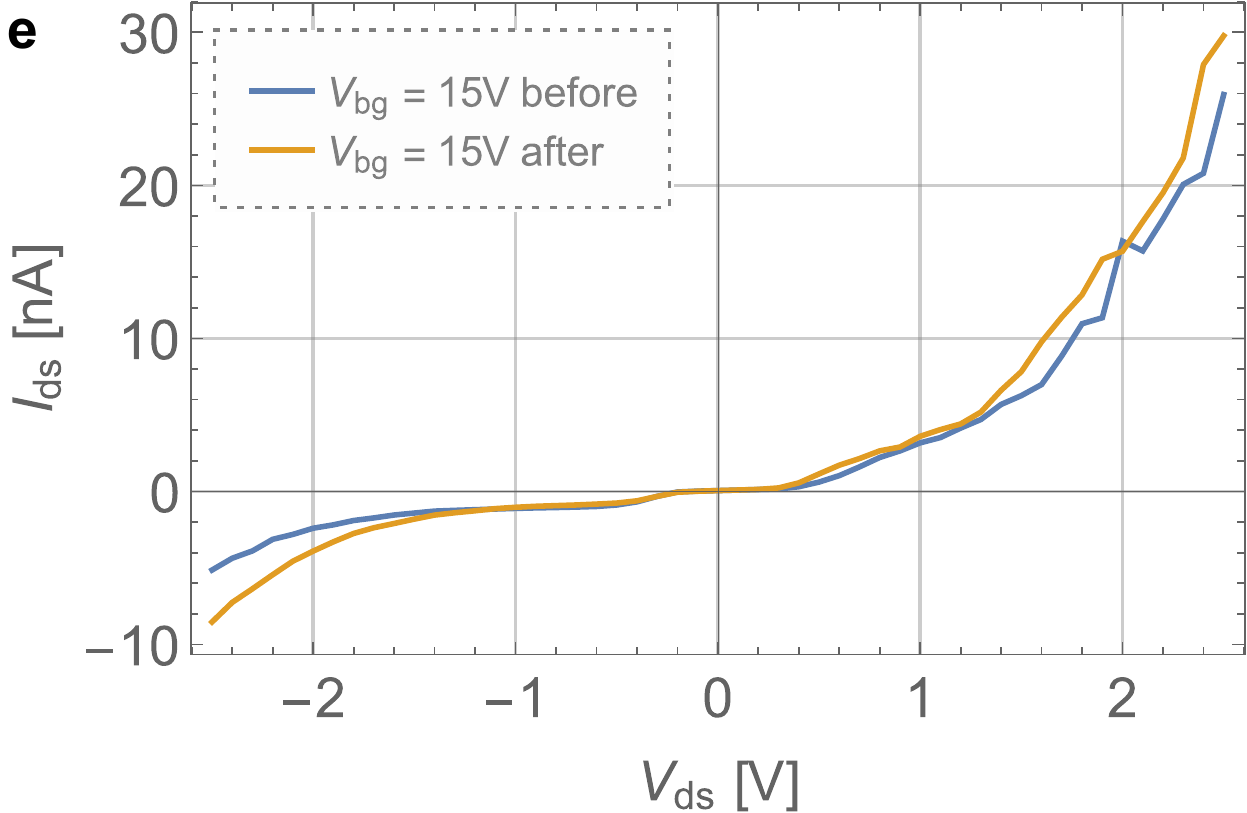}
\caption{\textbf{$\gamma$-ray tests of 2D material-based devices. (a)} PL spectra of a hBN quantum emitter before and after the $\gamma$-ray tests show no changes (vertically offset for clarity). \textbf{(b)} Second-order correlation function dipping at zero time delay to 0.185(23) before and to 0.188(25) after the irradiation. The values were obtained from fitting a three-level system. \textbf{(c)} Spectrum of a newly created quantum emitter after the $\gamma$-ray test. As the emitter was not annealed following the irradiation, its brightness and stability was lacking behind plasma-etched and annealed emitters. \textbf{(d)} Back gate sweeps before and after the irradiation with different drain-source biases $V_{\text{ds}}$. The orange dashed line was recorded 5$\,$hrs past the orange solid line to check for temporal variations. In terms of current ON/OFF ratio the temporal variations are larger than the variations caused by the $\gamma$-rays. Tuning the $V_{\text{ds}}$ can restore the initial performance (see area between green and red lines). \textbf{(e)} The $I$-$V$ curve at a fixed $V_{\text{bg}}=15\,$V shows only slight alteration after irradiation.}
\label{fig:3}
\end{figure*}

\begin{figure*}[t!]
\centering
  \includegraphics[width=0.67\linewidth,keepaspectratio,valign=t]{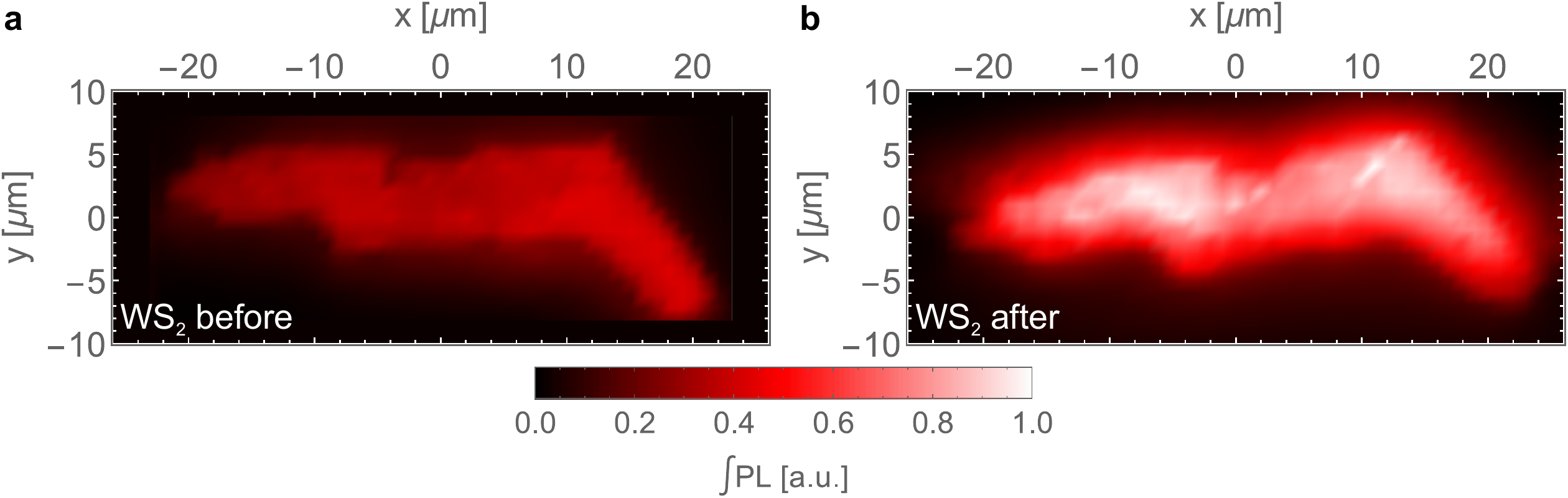}
  \includegraphics[width=0.32\linewidth,keepaspectratio,valign=t]{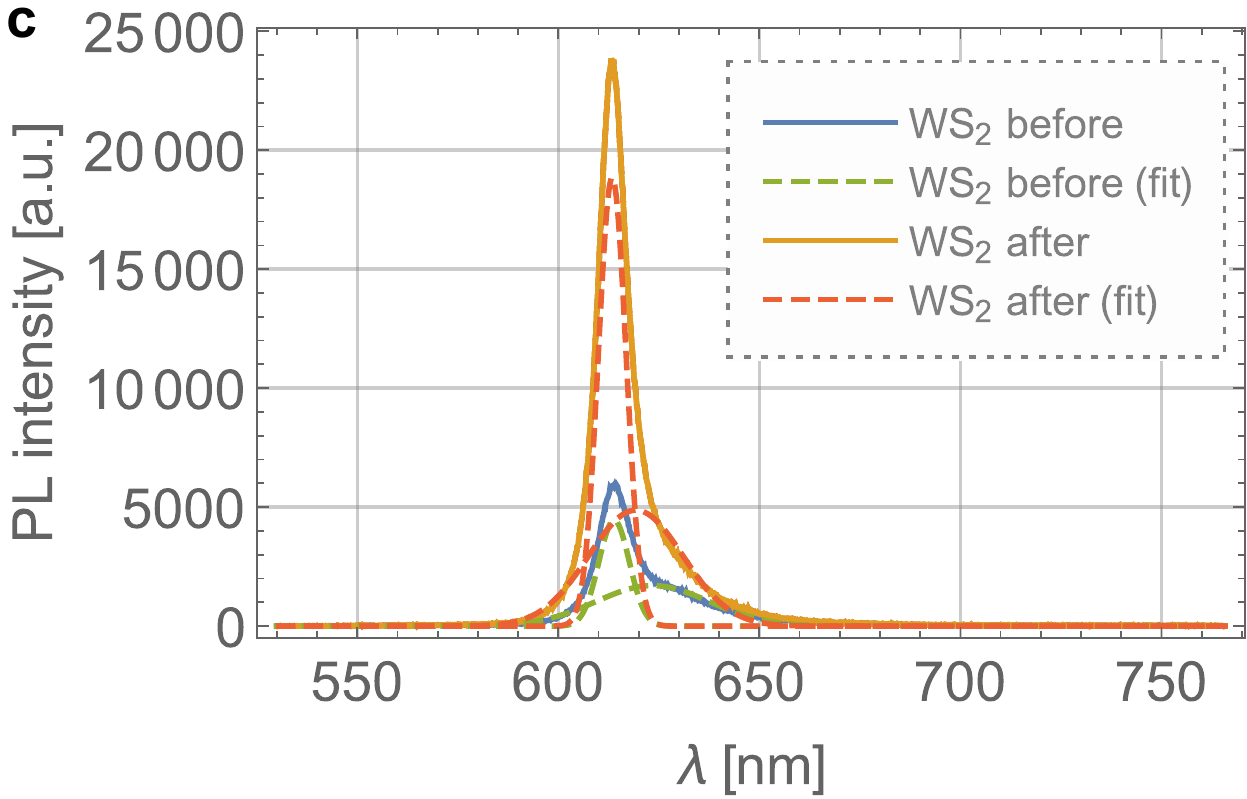}\\
  \vspace{0.1cm}
  \includegraphics[width=0.32\linewidth,keepaspectratio,valign=t]{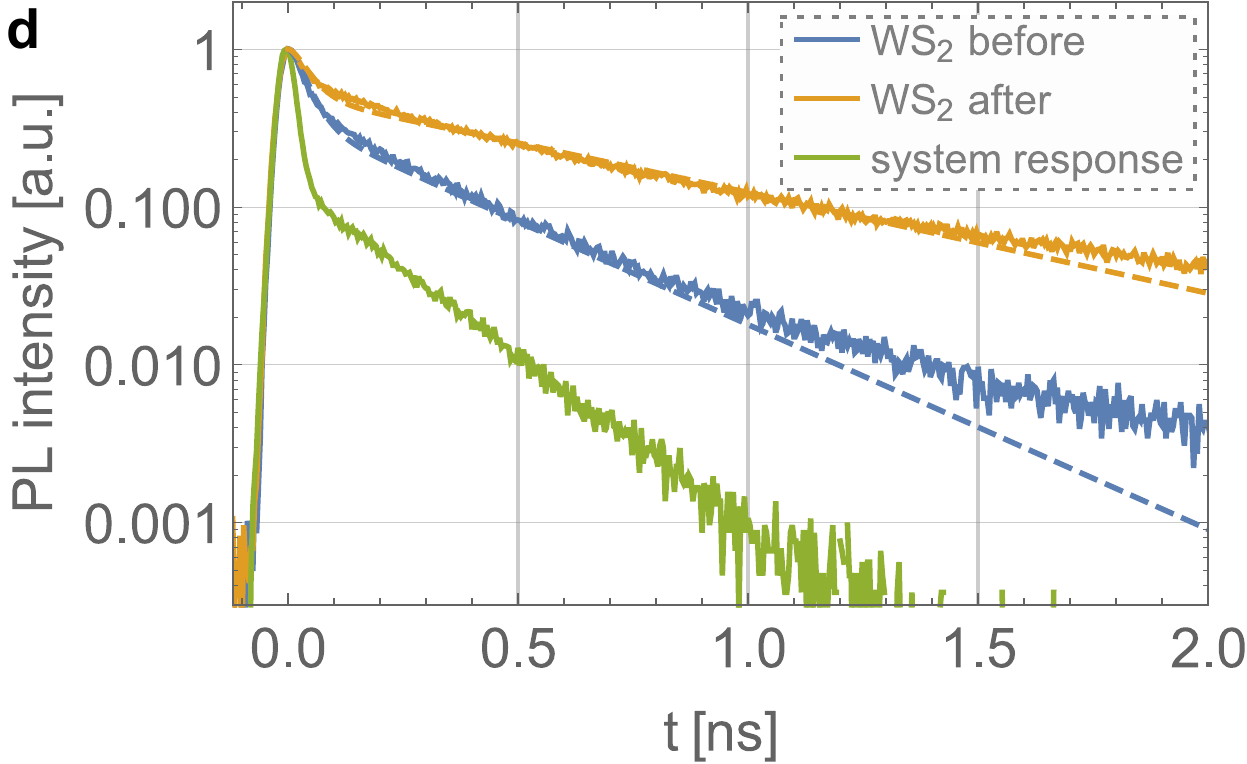}
  \includegraphics[width=0.32\linewidth,keepaspectratio,valign=t]{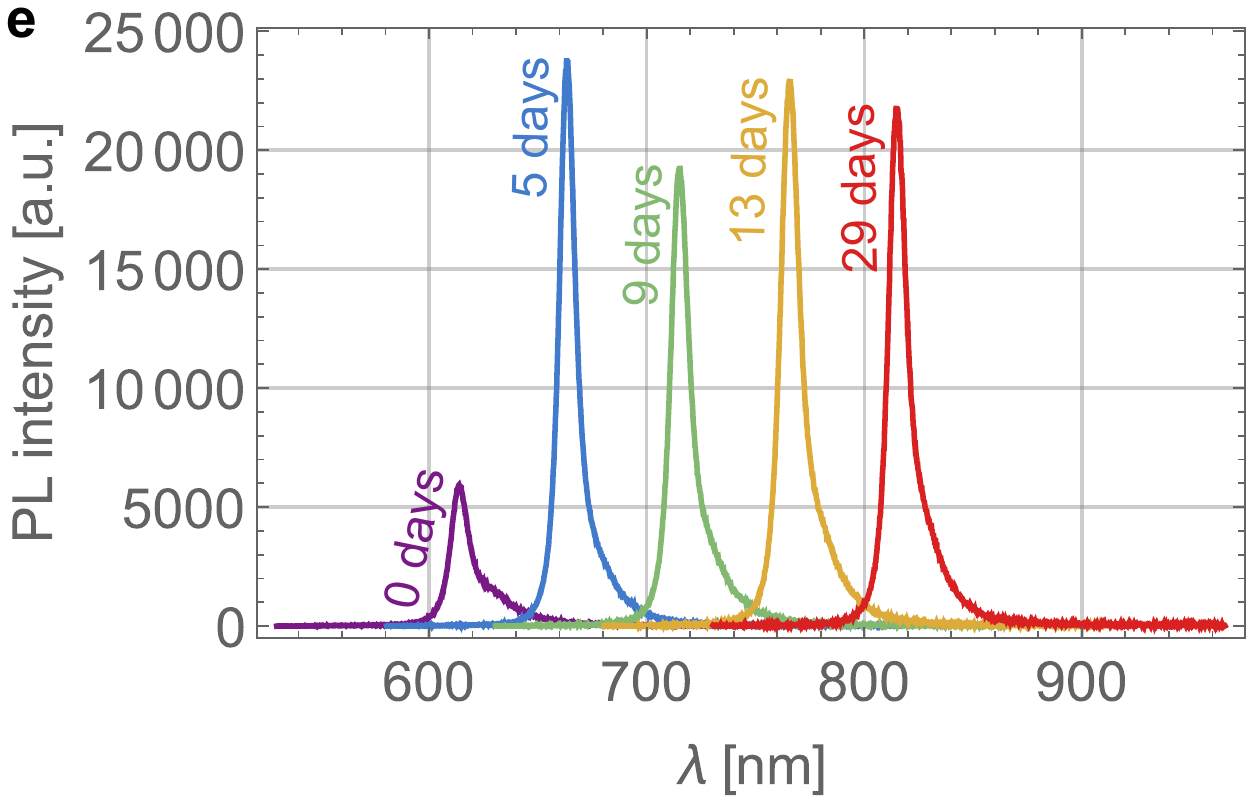}
  \includegraphics[width=0.32\linewidth,keepaspectratio,valign=t]{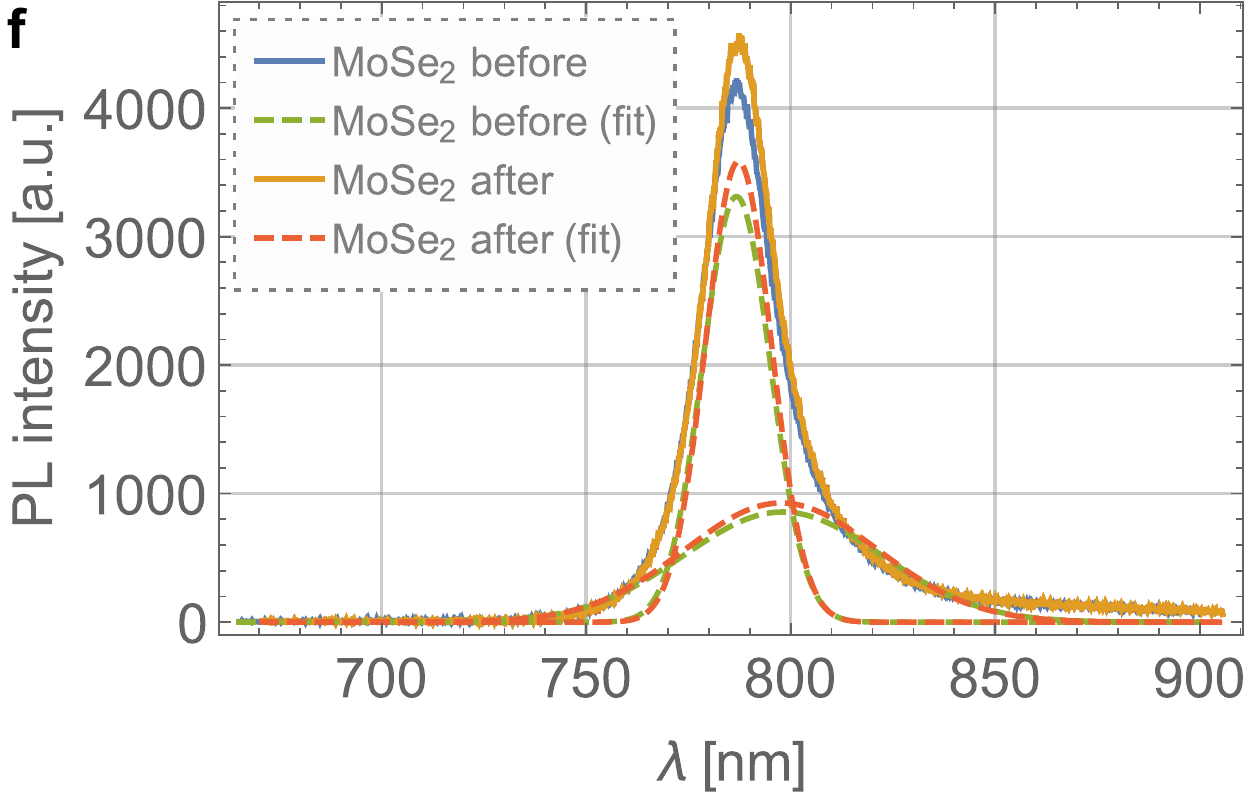}\\
  \vspace{0.1cm}
  \includegraphics[width=0.32\linewidth,keepaspectratio,valign=t]{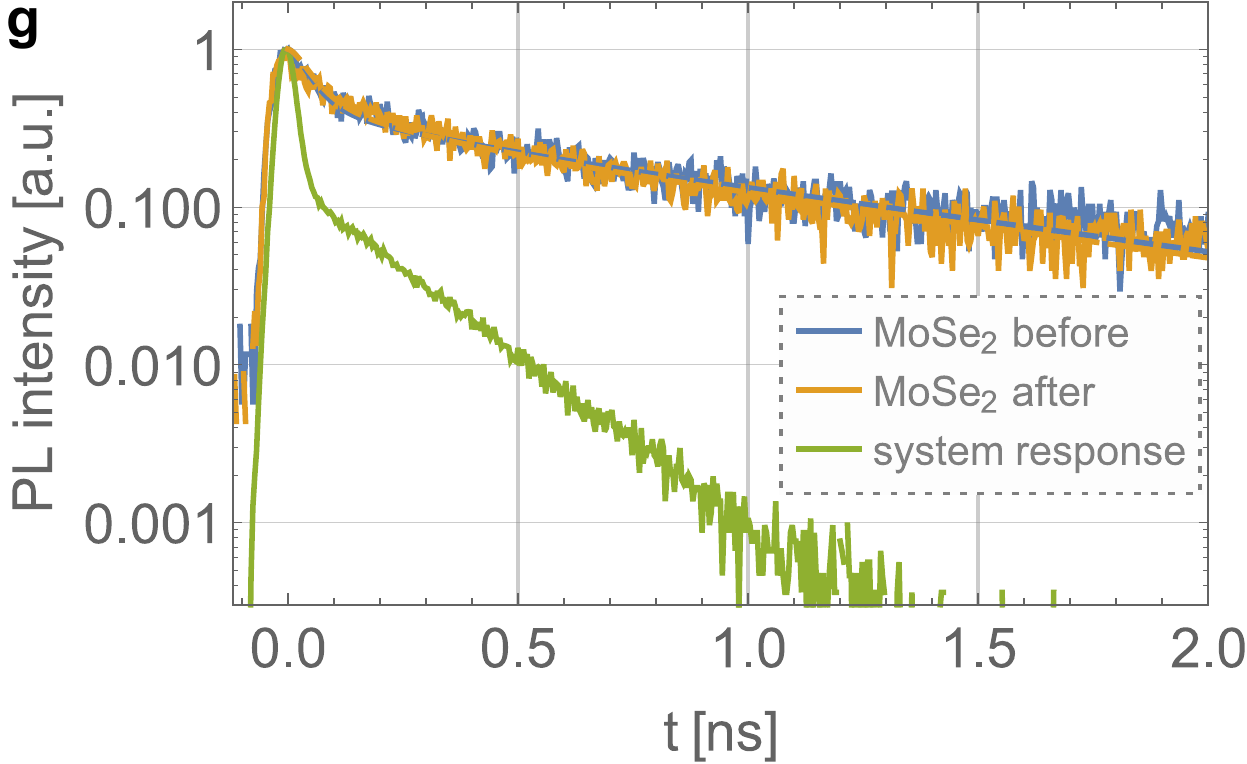}
  \includegraphics[width=0.32\linewidth,keepaspectratio,valign=t]{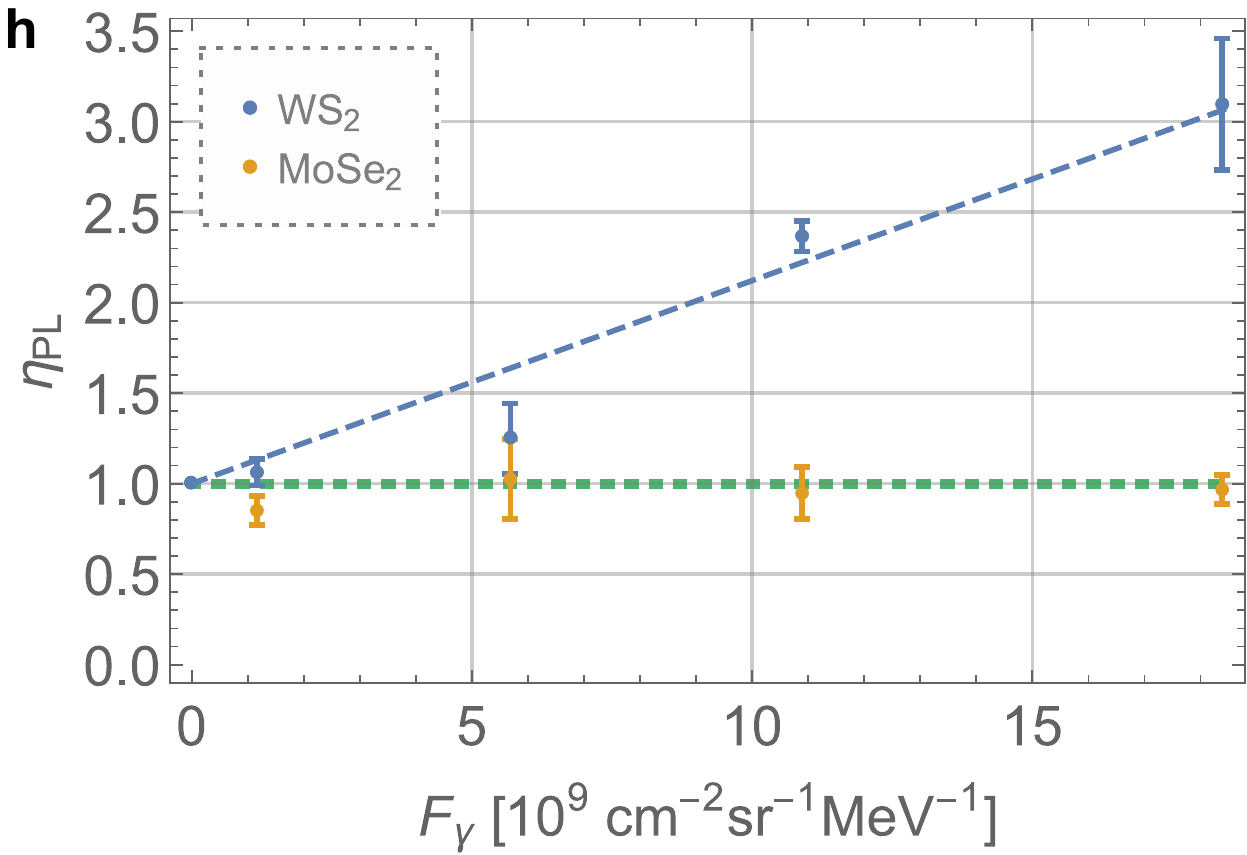}
  \includegraphics[width=0.32\linewidth,keepaspectratio,valign=t]{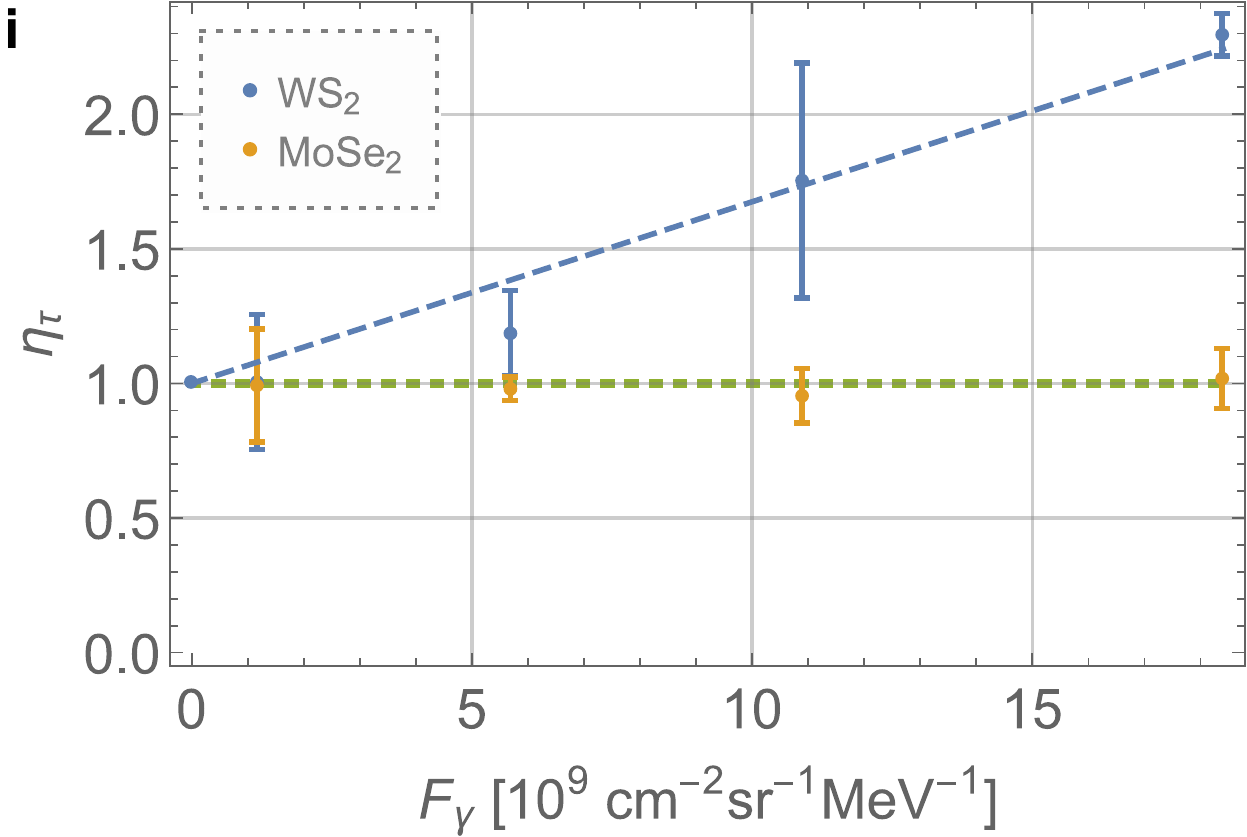}
\caption{\textbf{$\gamma$-ray tests of TMD monolayers. (a-e)} Optical characterization of a WS$_2$ monolayer before and after the $\gamma$-ray exposure. \textbf{(a,b)} The PL intensity maps scanned with $1\,\mu$m resolution integrated over the full spectrum show a strong increase in brightness after the irradiation. \textbf{(c)} In addition to the brightness increase of 2.99, the PL spectrum shows that the exciton/trion ratio also changed from 0.706(11) to 1.138(19). This was extracted from fitting two Gaussian distributions. \textbf{(d)} Similarly, the radiative carrier lifetime increased from 336(3) to 678(5)$\,$ps. The fit routine deconvolutes the data from the system response. \textbf{(e)} Long-term stability of the PL spectrum measured at different days. The irradiation took place at day 2. For clarity each subsequent spectra is shifted by 50$\,$nm. The peak wavelength remained invariant with its mean at 614.65$\,$nm. \textbf{(f,g)} PL emission spectrum and carrier lifetime for monolayer MoSe$_2$. This material remains predominantly unaffected by the gamma-rays, with the overall brightness increased by less than $5\%$ and the radiative lifetimes before and after the irradiation being 1086(41) and 1071(47)$\,$ps, respectively. Both sample monolayers experienced the same photon fluence. \textbf{(h,i)} Relative brightness and carrier lifetime increase as a function of $\gamma$-ray fluence averaged over all samples for WS$_2$ and MoSe$_2$. While there is little to no change for MoSe$_2$, for WS$_2$ the relative changes are linearly proportional to the radiation fluence. The data point at zero is the control sample. The green dashed line indicates $\eta=1$ (no change). The error bars are the standard deviation of the average.}
\label{fig:4}
\end{figure*}

The $\gamma$-ray source predominantly used for space qualification is the radioactive isotope $\ce{^{60}_{27}Co}$, which emits photons with energies of 1.17 and 1.33$\,$MeV as it decays. Due to availability, we used the isotope $\ce{^{22}_{11}Na}$ instead, which decays into $\ce{^{22}_{10}Ne}$ via the emission of a $1.28\,$MeV photon\cite{toi}, similar to the $\gamma$-ray energy from the Co isotope. With a branching ratio of approximately 9:1 the decay either happens via a $\beta^{+}$ transition or electron capture, respectively, resulting in a 90$\,\%$ probability that a positron is emitted. The positrons are shielded by Al foil, where they recombine with electrons to create two $\gamma$-rays with energies of 511$\,$keV in opposite directions. From its initial nominal activity of 1.04$\,$GBq, a total photon flux of 10.3$\,$MBq$\cdot$cm$^{-2}\cdot$sr$^{-1}\cdot$MeV$^{-1}$ is emitted into the output mode of the Tungsten container in which the source was kept. We placed the samples at distances of $d=10.0(1)$, 13.0(1), 18.0(1) and 40.0(1)$\,$cm to the source output, thus simulating various altitudes and times in orbit (see Supplementary Figure S1). All samples were irradiated for 2:27$\,$hrs, meaning that the maximal fluence at the closest distance to the source was F$_\gamma=18.41\times 10^9\cdot$cm$^{-2}\cdot$sr$^{-1}\cdot$MeV$^{-1}$. Unless stated otherwise, the crystals presented in this section were irradiated with the highest photon flux. For all experiments, unexposed control samples ensured that any potentially observed changes are solely due to irradiation.\\
\indent The performance of the single-photon emitters in hBN and the FET devices remained invariant when comparing samples before and after the $\gamma$-ray exposure. The zero phonon line (ZPL) of a sample quantum emitter as shown in Figure $\ref{fig:3}$(a) peaked initially at 563.78(8)$\,$nm with a linewidth of 4.29(13)$\,$nm (extracted from fit). Unless stated otherwise, all uncertainties are $95\%$ confidence intervals. After the crystal was irradiated, the ZPL peaked at 563.79(13)$\,$nm with a linewidth of 4.73(19)$\,$nm. Similarly, its $g^{(2)}(0)$ did not change (see Figure $\ref{fig:3}$(b)) with $g^{(2)}_{\text{i}}(0)=0.185(23)$ and $g^{(2)}_{\text{f}}(0)=0.188(25)$, where index i and f stand for before and after the exposure, respectively. While the quantum emitters already present in the hBN crystal did not change with respect to their optical emission properties, the $\gamma$-rays were able to create five new emitters on $\approx 40000\,\mu$m$^2$ of crystal area. Thus, the probability of creating a second emitter directly adjacent to another is very low. Figure $\ref{fig:3}$(c) shows the spectrum of one of the newly created emitters. As the crystal was not subsequently annealed, its brightness as well as stability was not as good as for other emitters\citep{10.1021/acsphotonics.8b00127}.\\
\indent Comparably, the FETs were also only marginally affected by the $\gamma$-rays. Figure $\ref{fig:3}$(d) shows back gate sweeps for a MoS$_2$ transistor. The current ON/OFF ratio $\beta$ was reduced from $\beta_\text{i}=21213$ to $\beta_\text{f}=14863$ at a drain-source bias of $V_{\text{ds}}=1\,$V. While this is a significant change in the ON/OFF ratio, we measured the ON/OFF ratio 5$\,$hrs later and saw $\beta$ further reduced to 4978 (dashed line in Figure $\ref{fig:3}$(d)). Hence, we attribute these changes in the ON/OFF ratio to temporal variations. The FETs in general are sensitive to surface adsorption which causes these temporal variations. However, by varying the drain-source bias from $0.8-1.2\,$V, the initial performance could be restored (see Figure $\ref{fig:3}$(d)). Another characteristic of transistors is the $I$-$V$ curve measured at fixed back gate voltages $V_{\text{bg}}$. For $V_{\text{bg}}=15\,$V this is shown in Figure $\ref{fig:3}$(e) and for other $V_{\text{bg}}$ in Supplementary Figure S2. The $I$-$V$ curves show no change due to the irradiation.\\
\indent While the 2D material-based devices showed no change after the $\gamma$-ray tests, the optical signature of monolayer WS$_2$ changed remarkably: The monolayer shown in Figure $\ref{fig:2}$(b) showed a significant increase in photoluminescence. Moreover, the brightness increased by a factor of 2.99 after being exposed to the $\gamma$-rays (see Figure $\ref{fig:4}$(a,b)). Furthermore, from the averaged PL spectrum of the monolayer (see Figure $\ref{fig:4}$(c), for details of the averaging algorithm see Supplementary Information S1), we extract that the exciton/trion ratio $\alpha$ changed from 0.706(11) to 1.138(19). Both, the exciton and trion emission were enhanced, however, the exciton emission was enhanced more strongly as the change from $\alpha<1$ to $\alpha>1$ shows. This also indicates a change in doping ratio. Given the initial linewidths of 3.80(3) for excitons and 14.07(13)$\,$nm for trions, there was no change in center wavelength of the exciton emission (613.89(3) to 613.41(2)$\,$nm) and only a slight change of the trion emission (623.45(19) to 619.28(19)$\,$nm). However, the linewidths changed to 3.45(2) for excitons and 11.73(13)$\,$nm for trions. In addition, the radiative carrier lifetime (see Figure $\ref{fig:4}$(d)) had also increased from 336(3) to 678(5)$\,$ps. The increase in PL and lifetime was persistent over a timeframe of a full month (see Figure$\ref{fig:4}$(e)). The small variations in the peak maxima are most likely due to laser defocusing, owing to the small Rayleigh length of the laser with the high NA objective (see Methods). Nevertheless, quantities independent of this, such as the exciton/trion ratio as well as carrier lifetime, remained fully stable at all measurement days. Moreover, other samples (see Supplementary Figure S9) were less affected by laser defocusing during the long-term stability tests. The stability could not be monitored any longer, as the samples have been subsequently irradiated with protons. We still note that the samples kept their increased photoluminescence during subsequent proton tests.\\
\indent Since free excitons easily scatter and recombine at trapped charge carriers at defect sites, a change in doping ratio as well as longer carrier lifetime and increased PL intensity likely indicates a reduction in defect density. By averaging over the full data set of samples at the corresponding distance to the source, it can be seen that the effect of an increased PL and lifetime is linearly proportional to the $\gamma$-ray flux (see Figure $\ref{fig:4}$(h,i), respectively). Interestingly, this effect was not observed for MoSe$_2$ monolayers, (see Figure $\ref{fig:4}$(f,i)). Moreover, under the same exposure conditions, the PL had only increased marginally by 1.05 compared to the 2.99 from the WS$_2$ sample presented previously. In addition, the exciton/trion ratio was stable with $\alpha_\text{i}=1.328(36)$ and $\alpha_\text{f}=1.317(36)$ as well as was the carrier lifetime with $\tau^\text{i}_\text{r}=1086(41)$ and $\tau^\text{f}_\text{r}=1071(47)$ (see Figure $\ref{fig:4}$(f,g)). It should be mentioned that MoSe$_2$ and WS$_2$ have intrinsically different exciton/trion ratios, since our MoSe$_2$ is a p-type and WS$_2$ is an n-type semiconductor. By averaging over all samples we found $\alpha_\text{MoSe2} = 1.252(86)$ and $\alpha_\text{WS2} = 0.715(117)$ (the uncertainty is the standard deviation).\\
\indent During the data analysis we noted that the $\gamma$-radiation dose was $24^2$ higher than intended. This was due to a calculation error. As a consequence, the highest $\gamma$-ray fluence is equivalent to 2170 years at 500$\,$km above the polar caps (see Table $\ref{table1}$), instead of the planned 4 years. However, in terms of space certification this is not an issue. If anything, this further confirms radiation resistance, as all crystals survived the higher radiation doses. Furthermore, this proves that 2D materials can withstand even harsher radiation environments than LEO, such as during solar flares or near nuclear reactors. For the WS$_2$ monolayer we extrapolate the effect of increased photoluminescence and carrier lifetime after 4 years in orbit to be less than $0.15\%$ and $0.10\%$, respectively.

\subsubsection*{Backtracing of healing mechanism}
\begin{figure*}[t!]
\centering
  \includegraphics[width=0.32\linewidth,keepaspectratio,valign=t]{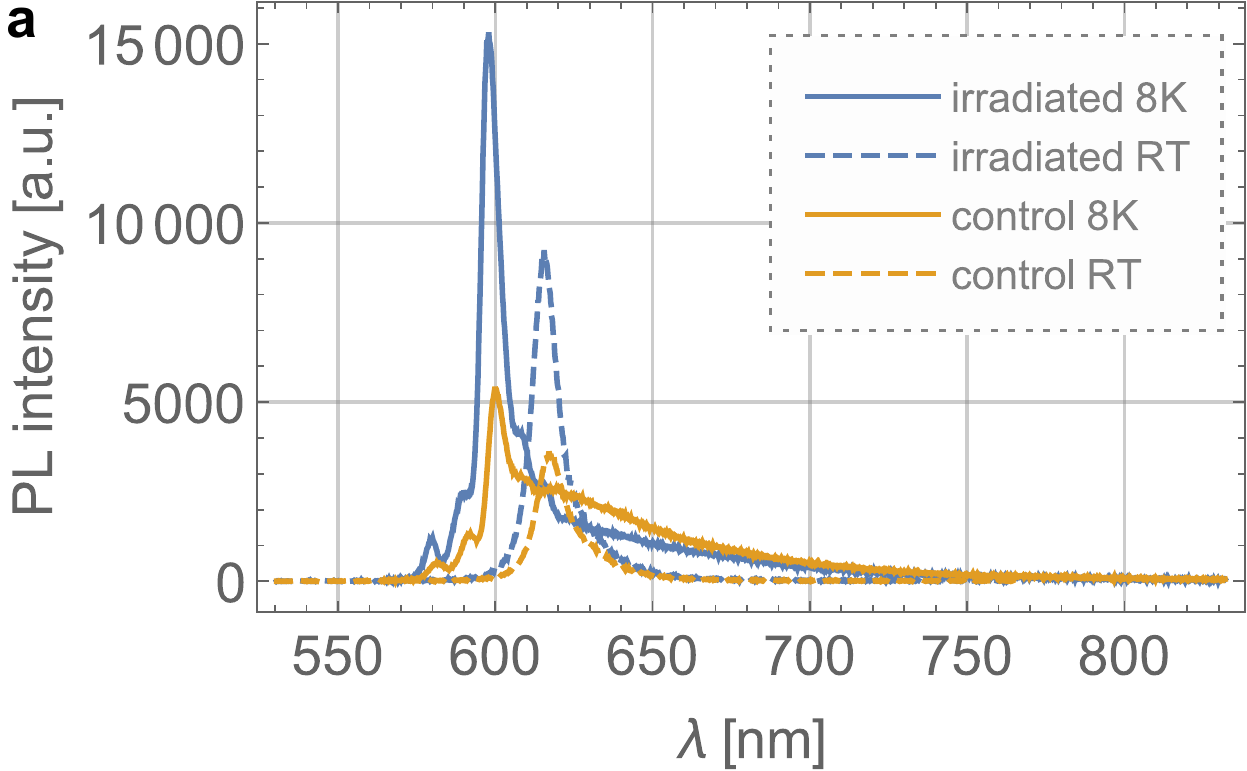}
  \includegraphics[width=0.32\linewidth,keepaspectratio,valign=t]{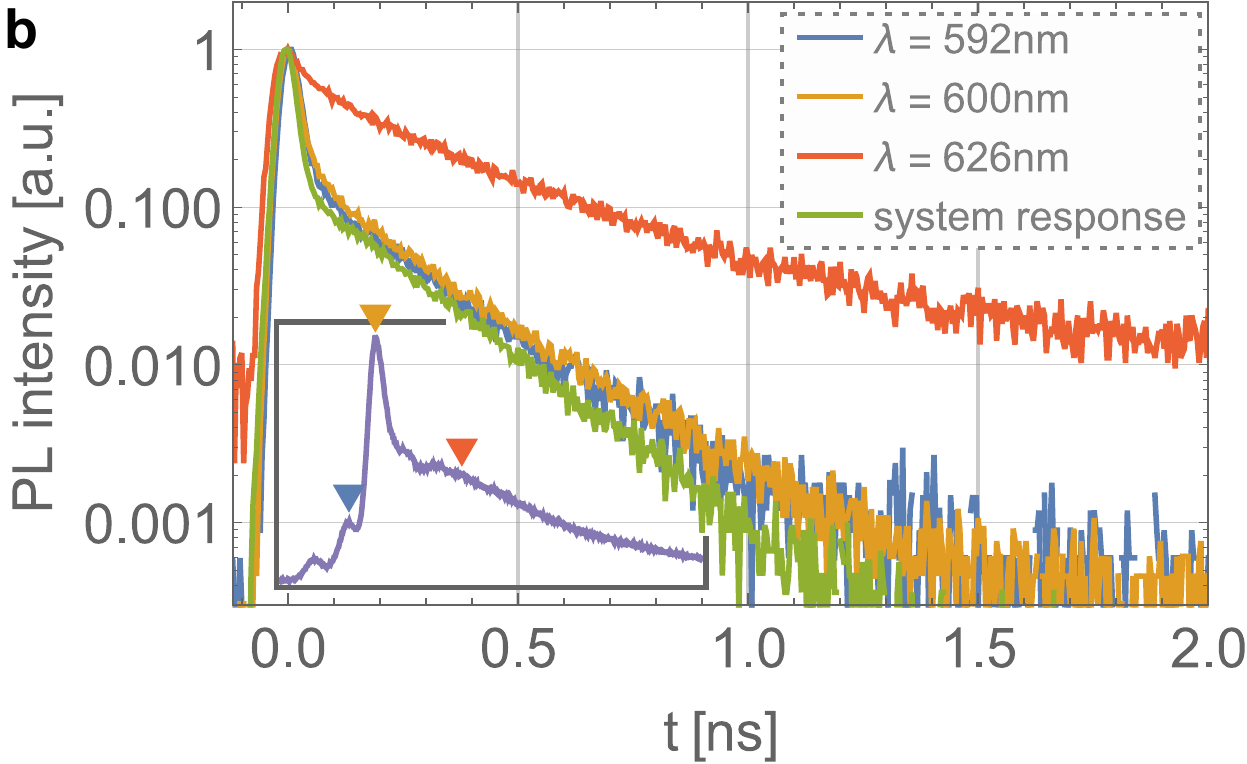}
  \includegraphics[width=0.32\linewidth,keepaspectratio,valign=t]{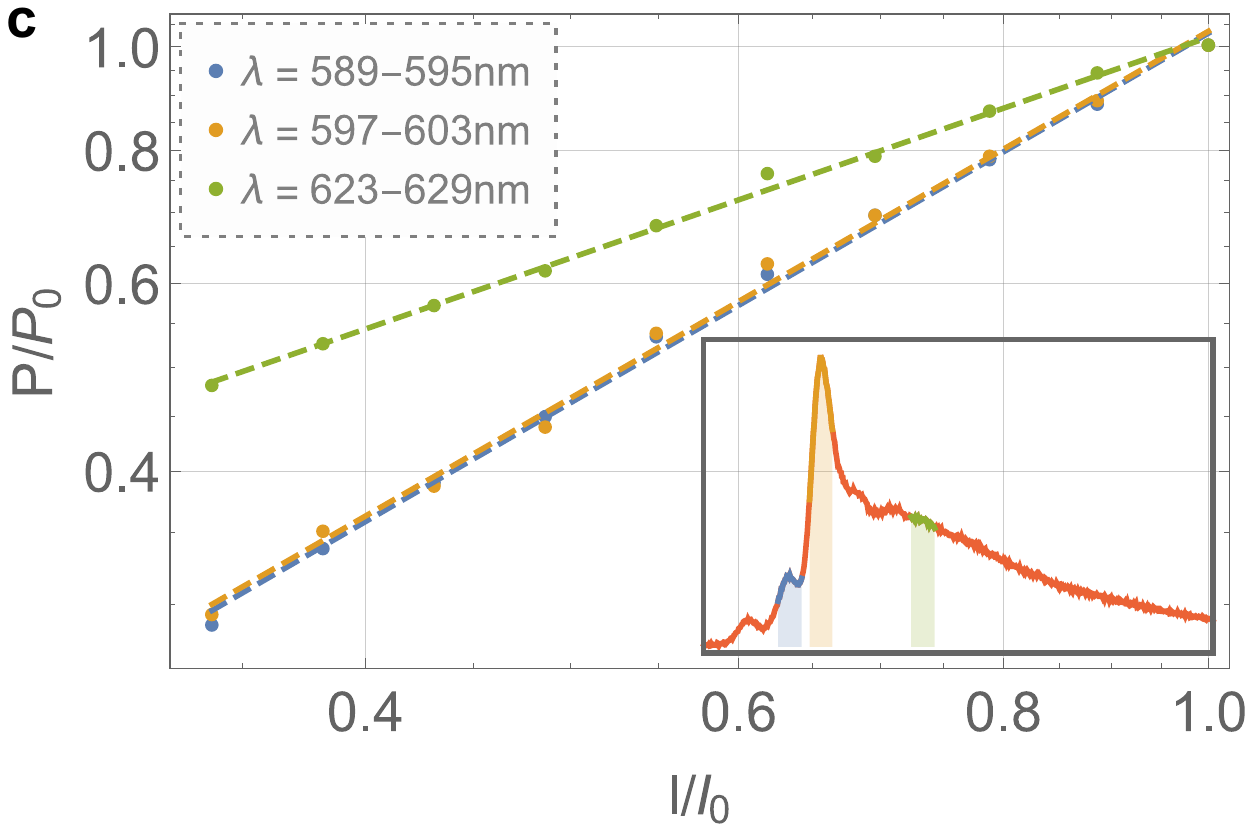}\\
  \vspace{0.1cm}
  \includegraphics[width=0.32\linewidth,keepaspectratio,valign=t]{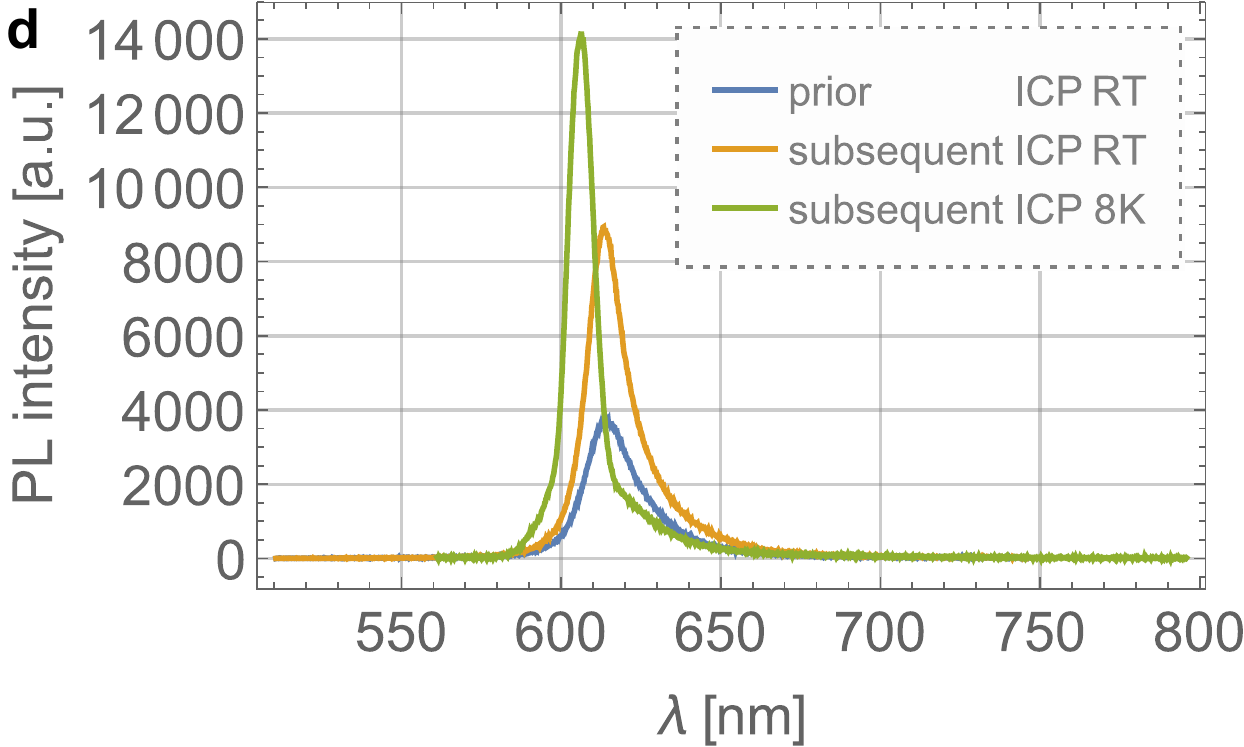}
  \includegraphics[width=0.32\linewidth,keepaspectratio,valign=t]{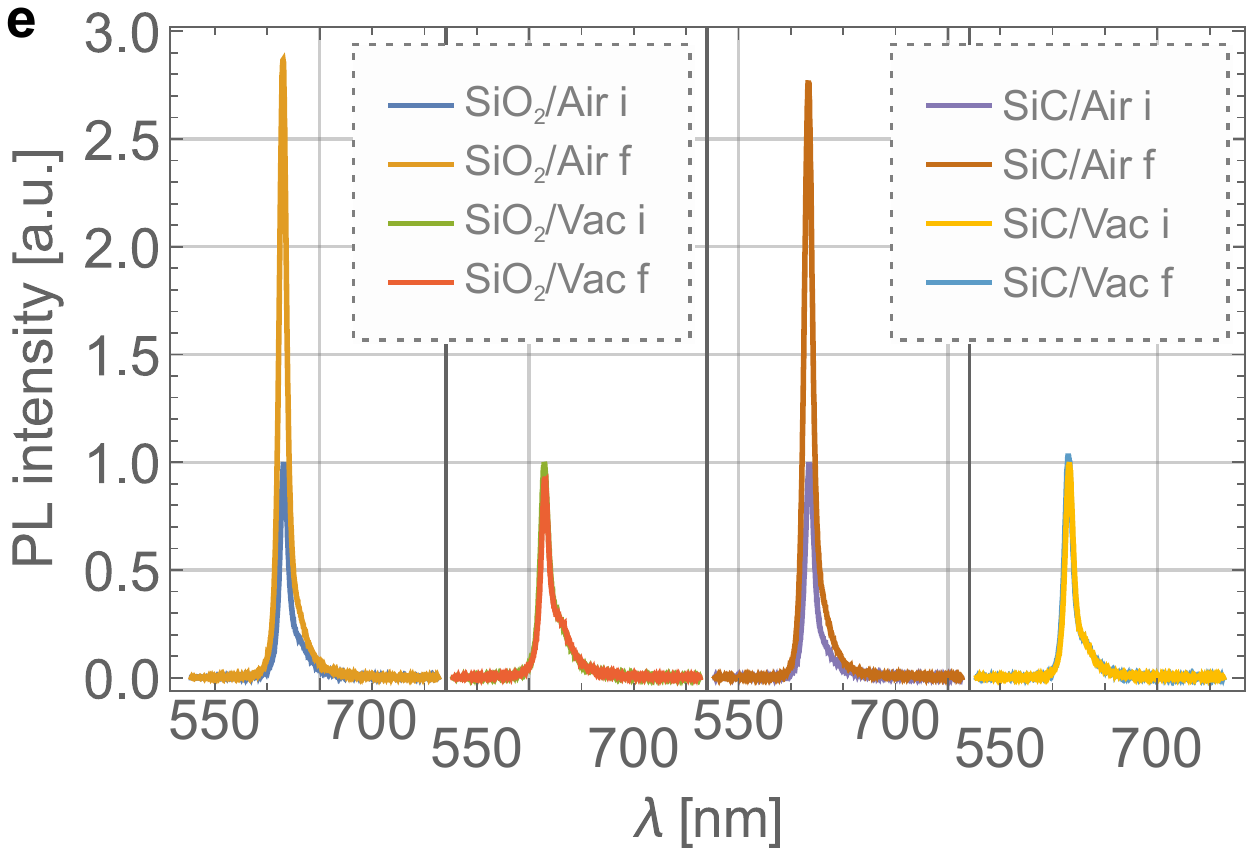}
  \includegraphics[width=0.32\linewidth,keepaspectratio,valign=t]{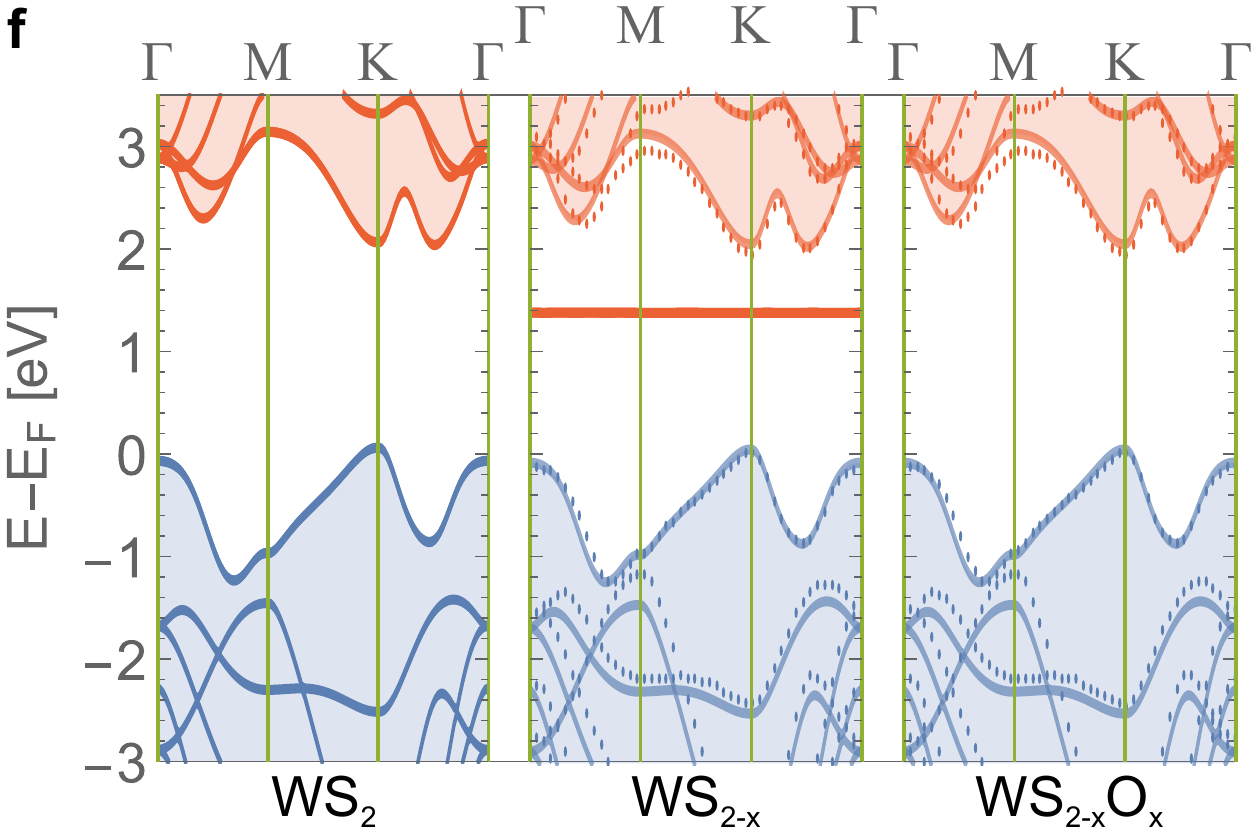}
\caption{\textbf{Identification of the $\gamma$-ray induced healing mechanism. (a)} PL spectrum of $\gamma$-irradiated and control sample at 8$\,$K and RT. The control sample shows strong defect emission in the red sideband. \textbf{(b)} Spectrally- and time-resolved PL reveals carrier lifetimes close to the system response at $\lambda=592,600\,$nm and 361(3)$\,$ps at 626$\,$nm, confirming excitonic and defect nature of the emission. Unlike for defects, the radiative lifetime for excitons/trions is proportional to temperature. The inset shows the positions in the spectrum (marked with triangles in the corresponding color) at which the lifetimes were measured. \textbf{(c)} Spectrally-resolved power dependence on a log-log-plot reveals a slope close to 1 at $\lambda=592,600\,$nm, indicating excitonic emission, while the slope $<1$ at 626$\,$nm means the emission originates from defects. The inset shows the positions in the spectrum (marked with the corresponding colors) at which the power dependence is measured. \textbf{(d)} PL emission prior and subsequent to the ICP treatment shows a similar increase in PL brightness compared to the $\gamma$-irradiated samples. In addition, at low temperature no defect emission becomes visible, confirming that oxygen can passivate vacancies. \textbf{(e)} Repetition of the $\gamma$-irradiation on SiO$_2$ and SiC substrates as well as in air and vacuum show that the atmosphere must be the source of oxygen used for the defect healing, most likely through adsorbed oxygen onto the surface. \textbf{(f)} DFT calculations of the bandstructure of pristine WS$_2$ (left), WS$_{2-x}$ (middle), WS$_{2-x}$O$_x$ (right) show that unlike the V$_\text{S}$ defect, the S$_\text{O}$ defect has no unoccupied deep mid-band gap state. The middle and right bandstructure show the conduction and valence band from the primitive pristine unit cell (solid lines) overlayed with the conduction and valence band from the supercell calculations (dotted lines).}
\label{fig:5}
\end{figure*}

An increase in photoluminescence and carrier lifetime is very surprising: Initially it was expected that radiation could lead to the formation of new defects, but not to defect healing. Although radiation-induced healing of nanomaterials has been reported\cite{10.1134/S2075113316020040}, such an effect has not been observed with $\gamma$-rays, specifically not with such remarkable consequences. It is known that the most common defects in exfoliated TMD materials are chalcogen vacancies\cite{10.1038/ncomms7293}. Simulations do also predict that these vacancies can chemically react with oxygen\cite{doi:10.1002/anie.201508828}, because oxygen itself is a chalcogen. Thus, we propose this healing is due to the chemadsorption of atmospheric oxygen, catalyzed by $\gamma$-irradiation. A similar mechanism was proposed in a study involving laser-induced defect healing of WSe$_2$\cite{doi:10.1021/acs.nanolett.5b00952}. The $\gamma$-ray induced healing observed in our study could happen via several different pathways. One possibility involves the formation of oxygen radicals due to the presence of free electrons from primary reactions like Compton scattering.\\
\indent In order to support this, we conducted low temperature PL measurements of irradiated and control samples at 8$\,$K. The low temperature environment reduces thermal broadening which allows the individual emission signature to be resolved. We discovered multiple blue-shifted peaks compared to the RT measurements (see Figure $\ref{fig:5}$(a)), most of which are attributed to negatively charged trions. Consistent with standard semiconductor models and experiments\cite{10.1364/AO.55.006251}, the exciton emission decreases with temperature as the trion emission increases. While both samples exhibit these low temperature excitonic features, the control sample showed additional PL emission in the red sideband. In contrast, the irradiated sample shows only weak emission in the red sideband. This becomes more evident by comparing the fraction of PL from trions and defects, which is 2.08:1 and 0.35:1 for irradiated and control sample, respectively. Therefore, the defect density had decreased 6-fold after the sample was exposed to the $\gamma$-radiation. We confirm this by measuring the spectrally- and time-resolved photoluminescence (TRPL): The PL emission is coupled to the single-photon counting module via a grating which makes the TRPL wavelength-selective. Figure $\ref{fig:5}$(b) shows the lifetime measurements for three wavelengths, with the positions marked with correspondingly-colored triangles in the spectrum (see inset). Unlike for defect states, the radiative lifetime of excitons and trions is directly proportional to the temperature\cite{PhysRevB.93.205423}. With the lifetime of the control sample at room temperature being 286(4)$\,$ps, we expect the lifetime of any excitonic emission at 8$\,$K to be around 7$\,$ps. In fact, at $\lambda=592$ and 600$\,$nm we measured a lifetime just above our system response time (which is $\approx 3\,$ps), much shorter than the room temperature measurements. However, at 626$\,$nm the radiative lifetime was 361(3)$\,$ps, thus indicating defect emission. Furthermore, we also measured the spectrally-resolved power dependence (see Figure $\ref{fig:5}$(c)) at the same wavelengths at which the TRPL was measured (marked with the corresponding colors in the inset). While the slope of the bands around 592 and 600$\,$nm are close to 1 if plotted on a log-log-scale, which means it originates from free excitons or trions, the slope at $626\,$nm is $<1$, which indicates defect emission\cite{10.1038/nphoton.2013.179}.\\
\indent In the next phase of this study we confirm that the defect healing is oxygen-related. We replicate the optical signature of the $\gamma$-ray exposed samples by treating freshly prepared monolayers with an O$_2$ inductively coupled plasma (ICP) and optimizing the plasma parameters (see Methods). Figure $\ref{fig:5}$(d) shows the PL spectrum prior and subsequent to the ICP treatment at RT and 8$\,$K. Much like the irradiated samples, also the monolayers treated with the oxygen plasma show a strong increase in brightness and no defect emission at low temperature as well as a longer carrier lifetime. While these results uphold the conjecture of oxygen-related defect healing, the oxygen could either be supplied by atmospheric oxygen or oxygen from the SiO$_2$ layer. Therefore, we repeated the irradiation with samples on both, Si/SiO$_2$ and SiC substrates in air and under vacuum at $10^{-4}\,$atm. This will also strongly reduce the amount of surface adsorbed oxygen. The in-air irradiated samples showed the characteristic increase in brightness and carrier lifetime, while the under vacuum irradiated samples remained unaffected (see Figure $\ref{fig:5}$(e)). We found no dependency on the substrate material. In the context of space certification, this means that WS$_2$ monolayers in evacuated spacecrafts are also not affected even by excessive $\gamma$-radiation. In addition to the WS$_2$ crystals, we also exposed MoS$_2$ and WSe$_2$ monolayers to gamma-rays (in air). We observed a slight increase in PL intensity and carrier lifetime after the irradiation for WSe$_2$ (see Supplementary Figure S8) and no change for MoS$_2$. It should be mentioned that the PL emission from monolayer MoS$_2$ in general is much weaker than for other TMDs, so any change is harder to observe. Furthermore, the change in PL and lifetime for the WSe$_2$ was much weaker than for WS$_2$, even though both samples experienced the same $\gamma$-ray fluence.\\
\indent We now turn to a theoretical analysis of the electronic structure of the proposed defect using density functional theory (DFT). Calculations were performed using the Vienna Ab initio Simulation Package (VASP)\cite{PhysRevB.54.11169,PhysRevB.59.1758}. The electronic bandstructure for pristine WS$_2$, WS$_{2-x}$ (with a sulfur vacancy: V$_\text{S}$) and WS$_{2-x}$O$_x$ (where an oxygen atom replaces a sulfur atom: S$_\text{O}$) are calculated using the GGA-PBE functional\cite{PhysRevLett.78.1396} (see Methods). The V$_\text{S}$ defect has a deep unoccupied state in the band gap (see Figure $\ref{fig:5}$(f), middle). This is consistent with our experimental observations: The sub-state is an acceptor state trapping electrons, which changes the doping ratio in the crystal. This means more charge carriers are available for charged excitons and thus the trion emission is enhanced. In turn, excitons recombine easily at defects leading to a shorter exciton lifetime. The S$_\text{O}$ defect shows no such deep defect state (see Figure $\ref{fig:5}$(f), right), meaning as soon as the vacancy is passivated with an oxygen atom, the electronic configuration is similar to the pristine crystal. Crystal structures of the proposed defects are shown in the supplementary information S3.\\
\indent The V$_\text{Se}$ and Se$_\text{O}$ defects in MoSe$_2$, however, show a very similar electronic structure (see Supplementary Figure S11), thus our DFT calculations alone cannot explain why the $\gamma$-ray induced defect healing only happens for WS$_2$. Selenide TMDs are known to have less chalcogen vacancies than sulfide TMDs. In fact, in our experiment we see this by the averaged longer carrier lifetime in MoSe$_2$ of 1264$\,$ps, revealing an inherently much smaller presence of defect sites. As already mentioned, scattering and recombination of excitons at defect sites leads to a reduced lifetime. This means intrinsically less defect healing can occur for MoSe$_2$. Furthermore, as the DFT calculations show, the defect state of the V$_\text{Se}$ is closer to the conduction band ($\Delta E =0.45\,$eV compared to $0.56\,$eV for the V$_\text{S}$ defect). This means that the non-radiative charge capture cross section (CCS) of the V$_\text{S}$ defect is smaller, as more phonons are required for the capture. For radiative charge capture this effect is reversed. With the energy difference of 0.45$\,$eV to the conduction band, the V$_\text{Se}$ defect has a smaller radiative CCS. The overall capture probability is given by defect density times capture cross section, so even though the non-radiative CCS for the V$_\text{Se}$ defect is higher, with the much-reduced defect density the overall capture probability is lower.

\subsubsection*{Proton and electron irradiation}
After the $\gamma$-ray tests, the samples were irradiated with high-energy charged particles, starting with protons and then electrons. In addition to the $\gamma$-irradiated samples, after each radiation test fresh samples were added to study both, combined and isolated radiation damage effects. The samples were irradiated with protons from a 1.7$\,$MV tandem accelerator. Due to the maximally available proton energy of 3.4$\,$MeV, the annual fluence spectra shown in Figure $\ref{fig:1}$(c) cannot directly be replicated. Instead, we integrate over the full spectrum for each orbital inclination, which yields 241.820, 721.318 and 464.770$\times 10^6\,$cm$^{-2}$ for 20$^\circ$, $51.6^\circ$ and $98^\circ$, respectively. Unfortunately, these fluence values are below the range of the used charge carrier counter, which is why we tested the samples at F$_{p^+}=10^{10}\,$cm$^{-2}$. However, at the lower proton energies (200, 500, 1000 and 2500$\,$keV) we used, the potential displacement damage caused by the protons\cite{10.1016/j.carbon.2010.12.057} is higher due to the higher stopping power of the 2D materials at lower energies (see Supplementary Information S4). As our fluence is anyway higher than required for 500$\,$km altitude, we cannot scale the fluence down according to the used proton energies. For all proton energies, we did not observe any changes in the device performances, PL spectra or carrier lifetimes (see Supplementary Information S5). Even after increasing the proton flux 100-fold, there were still no changes. This proton fluence of $10^{12}\,$cm$^{-2}$ corresponds to 1386$\,$years in orbit (at $51.6^\circ$ inclination and 500$\,$km altitude). Hence, we conclude that proton irradiation is no concern for 2D materials and devices in LEO.\\
\indent Finally, we exposed the samples to electrons using a scanning electron microscope (SEM). The damage to 2D materials caused by electrons is mostly displacement and sputtering\cite{10.1002/pssa.201670681}. Similar to the proton accelerator, both the energy range as well as the integrated flux are beyond the capabilities of the SEM. The tested energies were 5, 10, 20 and 30$\,$keV while the fluence varied from $10^{10}$ to $10^{15}\,$cm$^{-2}$. At 500$\,$km altitude, the integrated fluence are 12.35, 30.06 and 20.73$\times 10^6\,$cm$^{-2}$ for equatorial, ISS and polar orbit, respectively. At the lowest accessible fluence, which is still three orders of magnitude above what is expected in LEO, the crystals were mostly unaffected by the electron irradiation (see Supplementary Information S6). Extrapolating the fluence to LEO levels predict that electrons will not have any impact on 2D materials. Higher electron fluences result in permanent loss of photoluminescence for TMDs. We propose that this is due to the creation of chalcogen vacancies by knock-on damage, which, as previously mentioned, cause recombination, thereby quenching the PL\cite{PhysRevLett.109.035503}. However, if the electron energy is increased from 5 to 30$\,$keV, even at the highest fluence such damage was mitigated. This is because higher energy electrons have a smaller interaction cross section (see Supplementary Figure S12). The single-photon emitters remain unaffected by the electron irradiation, however, at extremely high fluences, the emitter density can be increased significantly\cite{doi:10.1021/acsami.6b09875,doi:10.1021/acsami.8b07506}. In our case this happened while focusing the SEM on a small crystal part before exposing the full crystals to the electrons. The experienced electron fluence at these positions was up to $10^{18}\,$cm$^{-2}$.

\section*{Discussion}
We presented a comprehensive study on the effects of radiation on 2D materials in vision of space certification. Moreover, this study covered the effects of \mbox{$\gamma$-}, proton and electron irradiation on TMD-based FETs and single-photon sources in hBN as well as their interaction with blank TMD monolayers. These nanomaterials were investigated back-to-back, shortly before and after irradiation. While all crystals remained effectively invariant under irradiation relevant for space environments, after excessive $\gamma$-radiation monolayer WS$_2$ exhibit significant increase in photoluminescence and carrier lifetime proportionally to the photon flux. This is attributed to the healing of sulfur vacancies induced by $\gamma$-radiation. We propose that the $\gamma$-rays, through a process like Compton scattering, dissociate atmospheric oxygen, which then chemically reacts with the vacancies. This mechanism was confirmed by low temperature measurements showing that defect emission was weakened upon $\gamma$-irradiation. Furthermore, bandstructure modeling of this reaction shows disappearing trapping sites, thus explaining the observed changes.\\
\indent A potential application of this effect could be a compact radiation dosimeter or radiation detector. In addition to the radiation tests, the low temperature measurements also confirm that 2D materials survive vacuum and thermal cycling. The tested radiation fluences were much higher than required for LEO. Hence, 2D materials and devices based on them have been proven to withstand the harsh space radiation. Moreover, 2D materials can even operate in environments with heavy irradiation, such as during solar flares or near nuclear reactors. Our results pave the way towards establishing the robustness and reliability of 2D material-based devices for space instrumentation. This combines the fields of space science and nanomaterials, thus opening new possibilities for future space missions.

\section*{Methods}
\subsection*{Device Fabrication.}
The bulk crystals were acquired from HQGraphene and used as received. After mechanical exfoliation onto Gel-Pak WF-40-X4, monolayer TMD and multilayer hBN crystals were optically identified and transferred via dry contact to Si/SiO$_2$ substrates (262$\,$nm thermally grown) or 4H-SiC substrates supplied by SiCrystal. The crystal thickness was confirmed using PSI measurements. The hBN crystals were exposed to an oxygen plasma generated from a microwave field at 200$\,$W for 1$\,$min and a pressure of 0.3$\,$mbar at a gas flow rate of 300$\,$cm$^3$/min at room temperature. The subsequent rapid thermal annealing was done under an Argon atmosphere at 850$\,^\circ$C at a gas flow of 500$\,$cm$^3$/min. The substrates for the FETs have been pre-patterned with gold electrodes using photolithography: After spin coating AZ MiR 701, the positive photoresist is exposed to UV light through a mask and developed. Using electron-beam thermal evaporation, 100$\,$nm of gold is deposited and then LOR 3A was used for lift-off. The monolayer crystals were mechanically transferred between the two electrodes with an approximate gap of 10$\,\mu$m, with an attached multilayer crystal touching the electrode completing the electrical connection. The two electrodes served as top gates (source and drain), while the heavily n$^+$-doped silicon substrate served as the back gate.
\subsection*{Optical Characterization.}
The home-built $\mu$PL setup utilized second harmonic generation to generate $522\,$nm ultrashort laser pulses (High Q Laser URDM). The linearly polarized laser is focused down to the diffraction limit by an Olympus 100$\times$/0.9 dry objective. For confocal PL mapping, the samples were moved on Newport precision stages with up to 0.2$\,\mu$m resolution. The in-reflection collected emission is wavelength filtered (Semrock RazorEdge ultrasteep long-pass edge filter), fully suppressing the pump light, while still collecting the full emission spectrum. This spectrum is recorded using a grating-based spectrometer (Princeton Instruments SpectraPro). The laser pulse length for time-resolved measurements is 300$\,$fs length at a repetition rate of 20.8$\,$MHz. The pulses were split into trigger signal and excitation pulse. The emitted photons were detected by a single-photon counter (Micro Photon Devices) after the grating, so that the time-resolved photoluminescence is also spectrally resolved. Both, trigger and single-photon signal were correlated by a PicoHarp 300. For low temperature measurements, a cryogenic chamber was added to the setup and the samples were cooled down to 8$\,$K with liquid He, at a pressure of 13$\,\mu$Torr to prevent the formation of ice on the window. The objective was replaced with a Nikon S Plan Fluor 60$\times$/0.7 objective with adjustable correction ring. The second-order correlation function was measured using a different confocal setup with a 512$\,$nm diode laser, equipped with a spectrometer and nano-positioning stage, ensuring that the defects can be localized. The correlation function data is fitted to a three-level system with ground and excited states as well as a meta-stable shelving state: 
\begin{align*}
g^{(2)}(\tau)=1-Ae^{-|\tau-\mu|/t_1}+Be^{-|\tau-\mu|/t_2}
\end{align*}
where $t_1$ and $t_2$ are the excited and meta-stable state lifetimes respectively, $\mu$ accounts for different electrical and optical path lengths in the HBT interferometer and $A$ and $B$ are the anti-bunching and bunching amplitudes. The experimental data has been normalized such that for very long time delays $g^{\left(2\right)}\left(\tau\rightarrow \infty\right)=1$.
\subsection*{Electrical Characterization.}
The FETs were characterized with a Kiethley 4200 Semiconductor Analyzer. One of the the gold electrodes is grounded, while the n$^+$-doped Si substrate functions as a back gate, providing uniform electrostatic doping in the monolayer. Back gate sweeps at different biases between source and drain were measured as well as $I$-$V$ curves at various back gate voltages. All electrical measurements were carried out at room temperature.
\subsection*{Irradiation.}
The radioactive isotope $\ce{^{22}_{11}Na}$ was used as a $\gamma$-ray source and was kept in a sealed Tungsten container, which was opened for the duration of the exposure. For every disintegration, a 1.275$\,$MeV photon is emitted into 4$\pi$. Due to a 90$\,\%$ chance of a positron being emitted, our experiments were shielded by thick Aluminum foil, thereby causing the positrons to recombine into two 511$\,$keV photons in opposite directions. The nominal activity was 1.04$\,$GBq, which together with the container geometry leads to a total photon flux of 10.3$\,$MBq$\cdot$cm$^{-2}\cdot$sr$^{-1}\cdot$MeV$^{-1}$ at the output of the container. The samples were placed at different distances to the source, simulating different altitudes/times in orbit, with a placement accuracy of 1$\,$mm. All samples were mounted facing towards the $\gamma$-ray source. The second $\gamma$-ray experiment took place 117 days later, after which the source activity decreased to 91.8$\%$ ($\tau_{1/2} = 2.603\,$years). We accounted for this by adjusting the distance to the source. We exposed samples in air and in a vacuum chamber at $10^{-4}\,$atm. The $\gamma$-rays were attenuated by the glass window ports of the vacuum chamber by only $5\%$. This attenuation does not account for the complete disappearance of the healing effect on the samples in the chamber. For the proton irradiation a high-energy implanter featuring a 1.7$\,$MV tandem Pelletron accelerator was used. TiH was used as target for the ion sputter source and Ti ions were filtered by a $90^\circ$ magnet. The tandem accelerator can double maximal proton energy, however, due to the used configuration the proton energy was limited from 200$\,$keV to 2.5$\,$MeV. The ion energy is typically well defined within $\pm 5\,$keV and the error on the fluence is less than $\pm 10\%$. The irradiation took place under pressures of $10^{-7}\,$Torr at room temperature. For the electron irradiation the SEM from an FEI Helios 600 NanoLab was used, allowing for electron energies ranging from 1 to 30$\,$keV at 2.2$\,$mPa and room temperature. The current was varied from 0.17 to 0.69$\,$nA. The electron fluence $F_e$ is given by $F_e=\frac{I\cdot t}{e\cdot A}$ where $I$ is the electron current, $t$ the frame time, $e$ the electron charge and $A$ the frame area. The crystal flakes were located at a very low electron flux and then the SEM was aligned using another flake nearby, so that the crystal flake under investigation is targeted with a focused electron beam.
\subsection*{Plasma etching.}
We used the commercial ICP-RIE system Samco RIE-400iP and varied all process parameters. We found the optimal process parameters to be 75$\,$W ICP power, 0$\,$W RF power, 3$\,$min plasma interaction time as well as a gas pressure of 6.6$\,$Pa at an oxygen gas flow rate of 30$\,$cm$^3$/min. The RF power is chosen zero to avoid any ion bombardment during the plasma exposure, thus ensuring the process is chemical and not physical (this results in crystal etching or thinning). All ICP processes were carried out at room temperature.
\subsection*{Computational Methods.}
The space environment calculations were performed using the SPENVIS web interface. The proton and electron flux spectra were calculated using the AP-8 MAX and AE-8 MAX models. The shielded fluence spectra for 1.853$\,$mm Al shielding were obtained using the MFLUX package. The interactions between charge carriers and matter are calculated using Monte Carlo simulations (see Supplementary Information S4)\cite{10.1016/j.nimb.2010.02.091,ESTAR,doi:10.1002/sca.20000}. These simulations take electromagnetic scattering processes and hadronic nuclear interactions into account. Due to the more complicated nature of the interactions of electrons with the shielding material, the electron fluence spectra are less accurate. The DFT calculations have been performed using the ab-initio total-energy and molecular-dynamics program VASP (Vienna Ab initio Simulation Package) developed at the Fakult\"at f\"ur Physik of the Universit\"at Wien\cite{PhysRevB.54.11169,PhysRevB.59.1758}. First, the geometry of the pristine conventional cell was optimized using a $15\times 15\times 1$ Monkhorst-Pack reciprocal space grid such that all forces were less than 0.001$\,$eV/\AA. We used a plane-wave energy cutoff of 450$\,$eV and norm-conserving pseudopotentials with nonlinear core-correction to describe the core electrons. We also used the Perdew-Burke-Ernzerhof (PBE) functional to describe the exchange-correlation energy. The monolayer was constructed using a $7\times 7\times 1$ supercell of the optimized primitive unit cell. The ionic positions were then relaxed again, while keeping the cell size fixed. We chose the vacuum distance between each layer, described by the lattice parameter $c$, such that the bandstructure is flat in $\Gamma$ to A direction of the Brillouin zone. This indicates that there is no inter-layer interaction. We used the same method to obtain the bandstructure of the oxygen and vacancy centers in both, WS$_2$ and MoSe$_2$. These calculations show flat bands in each high symmetry direction, which indicates that there is minimal defect-defect interaction between neighboring supercells. The effective bandstructures shown here were unfolded using the PyVaspwfc package\cite{PhysRevB.85.085201,PyVaspwfc}.\\
\indent When analyzing these calculations, it is important to remember that PBE DFT systematically underestimates the quasiparticle bandgap\cite{PhysRevLett.51.1884}. Further, verifying the DFT bandgap against the experimental optical bandgap requires consideration of the exciton binding energy, which is significant in 2D TMDs ($\sim 1\,$eV)\cite{10.1088/2053-1583/4/1/015026}. Noting these problems, we only consider our calculations as accurate enough to qualitatively predict the presence and relative ordering of unoccupied defect levels in the bandgap. To confirm our conclusions, future calculations should apply GW corrections.

\section*{Data availability}
The data that supports the findings of this study are available from the corresponding author upon request.

\begin{acknowledgments}
This work was funded by the Australian Research Council (CE110001027, FL150100019, DE140100805, DE170100169, DE160100098 and DP180103238). We acknowledge financial support from ANU PhD scholarships, the China Scholarship Council and the ANU Major Equipment Committee fund (No. 14MEC34). We thank the ACT Node of the Australian National Fabrication Facility for access to their nano- and microfabrication facilities. This research was undertaken with the assistance of resources and services from the National Computational Infrastructure (NCI). We also thank Hark Hoe Tan for access to the TRPL system. We acknowledge access to NCRIS facilities (ANFF and the Heavy Ion Accelerator Capability) at the Australian National University, in particular we thank Rob Elliman and Tom Ratcliff for assistance with the implanter.
\end{acknowledgments}

\section*{Additional information}
\subsection*{Author contributions}
P.K.L., Y.L. and B.C.B. supervised the project. T.V. and P.K.L. devised the experiments. T.V., K.S., A.S. and L.Z. fabricated and characterized the samples. T.V., P.R. and M.W.D. conducted the theoretical calculations. T.V., J.S. and J.R.M. carried out the irradiation experiments. T.V. and F.K. developed the ICP processes. All authors participated in the results discussions, analysis and the writing of the manuscript.
\subsection*{Supplementary information}
Supplementary Information accompanies this paper at URL.
\subsection*{Competing interests}
The authors declare no competing financial interests.
\subsection*{Materials \& Correspondence}
Correspondence and requests for materials should be addressed to T.V. (email: tobias.vogl@anu.edu.au) or to P.K.L. (email: ping.lam@anu.edu.au).

\begin{table}[b]
\caption{\label{table1}Summary of total measured gamma-ray flux from the CORONAS-I satellite at 500$\,$km altitude. Full data available in the reference\cite{bucik1999}.}
\begin{ruledtabular}
\begin{tabular}{ccc}
    Location & $\gamma$-ray energy [MeV] & Flux [cm$^{-2}$s$^{-1}$sr$^{-1}$MeV$^{-1}$]\\
\colrule
    Equator & 0.32 - 1 & 0.079(59)\\
    Equator & 1 - 3 & 0.022(14)\\
    Polar cap & 0.32 - 1 & 0.174(59)\\
    Polar cap & 1 - 3 & 0.095(14)\\
\end{tabular}
\end{ruledtabular}
\end{table}

\clearpage

\begin{thebibliography}{10}
\expandafter\ifx\csname url\endcsname\relax
  \def\url#1{\texttt{#1}}\fi
\expandafter\ifx\csname urlprefix\endcsname\relax\def\urlprefix{URL }\fi
\providecommand{\bibinfo}[2]{#2}
\providecommand{\eprint}[2][]{\url{#2}}

\bibitem{10.1038/nnano.2011.56}
\bibinfo{author}{Cheng, G.} \emph{et~al.}
\newblock \bibinfo{title}{Sketched oxide single-electron transistor}.
\newblock \emph{\bibinfo{journal}{Nat. Nanotech.}}
  \textbf{\bibinfo{volume}{6}}, \bibinfo{pages}{343--347}
  (\bibinfo{year}{2011}).
\newblock \urlprefix\url{https://www.nature.com/articles/nnano.2011.56}.

\bibitem{10.1038/nnano.2012.21}
\bibinfo{author}{Fuechsle, M.} \emph{et~al.}
\newblock \bibinfo{title}{A single-atom transistor}.
\newblock \emph{\bibinfo{journal}{Nat. Nanotech.}}
  \textbf{\bibinfo{volume}{7}}, \bibinfo{pages}{242--246}
  (\bibinfo{year}{2012}).
\newblock \urlprefix\url{https://www.nature.com/articles/nnano.2012.21}.

\bibitem{10.1126/science.1102896}
\bibinfo{author}{Novoselov, K.~S.} \emph{et~al.}
\newblock \bibinfo{title}{Electric field effect in atomically thin carbon
  films}.
\newblock \emph{\bibinfo{journal}{Science}} \textbf{\bibinfo{volume}{306}},
  \bibinfo{pages}{666--669} (\bibinfo{year}{2004}).
\newblock \urlprefix\url{http://science.sciencemag.org/content/306/5696/666}.

\bibitem{PhysRevLett.105.136805}
\bibinfo{author}{Mak, K.~F.}, \bibinfo{author}{Lee, C.}, \bibinfo{author}{Hone,
  J.}, \bibinfo{author}{Shan, J.} \& \bibinfo{author}{Heinz, T.~F.}
\newblock \bibinfo{title}{Atomically thin ${\mathrm{mos}}_{2}$: A new
  direct-gap semiconductor}.
\newblock \emph{\bibinfo{journal}{Phys. Rev. Lett.}}
  \textbf{\bibinfo{volume}{105}}, \bibinfo{pages}{136805}
  (\bibinfo{year}{2010}).
\newblock
  \urlprefix\url{https://link.aps.org/doi/10.1103/PhysRevLett.105.136805}.

\bibitem{10.1038/nnano.2010.89}
\bibinfo{author}{Schwierz, F.}
\newblock \bibinfo{title}{Graphene transistors}.
\newblock \emph{\bibinfo{journal}{Nat. Nanotechnol.}}
  \textbf{\bibinfo{volume}{5}}, \bibinfo{pages}{487--496}
  (\bibinfo{year}{2010}).
\newblock \urlprefix\url{http://dx.doi.org/10.1038/nnano.2010.89}.

\bibitem{10.1038/nnano.2010.279}
\bibinfo{author}{Radisavljevic, B.}, \bibinfo{author}{Radenovic, A.},
  \bibinfo{author}{Brivio, J.}, \bibinfo{author}{Giacometti, V.} \&
  \bibinfo{author}{Kis, A.}
\newblock \bibinfo{title}{Single-layer mos$_2$ transistors}.
\newblock \emph{\bibinfo{journal}{Nat. Nanotech.}}
  \textbf{\bibinfo{volume}{6}}, \bibinfo{pages}{147--150}
  (\bibinfo{year}{2011}).
\newblock \urlprefix\url{https://www.nature.com/articles/nnano.2010.279}.

\bibitem{doi:10.1021/nn501723y}
\bibinfo{author}{Roy, T.} \emph{et~al.}
\newblock \bibinfo{title}{Field-effect transistors built from all
  two-dimensional material components}.
\newblock \emph{\bibinfo{journal}{ACS Nano}} \textbf{\bibinfo{volume}{8}},
  \bibinfo{pages}{6259--6264} (\bibinfo{year}{2014}).
\newblock \urlprefix\url{https://doi.org/10.1021/nn501723y}.

\bibitem{10.1038/nnano.2016.115}
\bibinfo{author}{Zhao, M.} \emph{et~al.}
\newblock \bibinfo{title}{Large-scale chemical assembly of atomically thin
  transistors and circuits}.
\newblock \emph{\bibinfo{journal}{Nat. Nanotech.}}
  \textbf{\bibinfo{volume}{11}}, \bibinfo{pages}{954--959}
  (\bibinfo{year}{2016}).
\newblock \urlprefix\url{https://www.nature.com/articles/nnano.2016.115}.

\bibitem{10.1126/science.aah4698}
\bibinfo{author}{Desai, S.~B.} \emph{et~al.}
\newblock \bibinfo{title}{Mos$_2$ transistors with 1-nanometer gate lengths}.
\newblock \emph{\bibinfo{journal}{Science}} \textbf{\bibinfo{volume}{354}},
  \bibinfo{pages}{99--102} (\bibinfo{year}{2016}).
\newblock \urlprefix\url{http://science.sciencemag.org/content/354/6308/99}.

\bibitem{10.1088/1674-4926/38/3/031002}
\bibinfo{author}{Huo, N.}, \bibinfo{author}{Yang, Y.} \& \bibinfo{author}{Li,
  J.}
\newblock \bibinfo{title}{Optoelectronics based on 2d tmds and
  heterostructures}.
\newblock \emph{\bibinfo{journal}{J. Semicond.}} \textbf{\bibinfo{volume}{38}},
  \bibinfo{pages}{031002} (\bibinfo{year}{2017}).
\newblock \urlprefix\url{http://stacks.iop.org/1674-4926/38/i=3/a=031002}.

\bibitem{10.1038/s41699-018-0049-3}
\bibinfo{author}{Furchi, M.~M.} \emph{et~al.}
\newblock \bibinfo{title}{Device physics of van der waals heterojunction solar
  cells}.
\newblock \emph{\bibinfo{journal}{npj 2D Materials and Applications}}
  \textbf{\bibinfo{volume}{2}}, \bibinfo{pages}{3} (\bibinfo{year}{2018}).
\newblock \urlprefix\url{https://www.nature.com/articles/s41699-018-0049-3}.

\bibitem{10.1039/C6TA09831B}
\bibinfo{author}{Shi, L.} \& \bibinfo{author}{Zhao, T.}
\newblock \bibinfo{title}{Recent advances in inorganic 2d materials and their
  applications in lithium and sodium batteries}.
\newblock \emph{\bibinfo{journal}{J. Mater. Chem. A}}
  \textbf{\bibinfo{volume}{5}}, \bibinfo{pages}{3735--3758}
  (\bibinfo{year}{2017}).
\newblock \urlprefix\url{http://dx.doi.org/10.1039/C6TA09831B}.

\bibitem{nnano.2015.242}
\bibinfo{author}{Tran, T.~T.}, \bibinfo{author}{Bray, K.},
  \bibinfo{author}{Ford, M.~J.}, \bibinfo{author}{Toth, M.} \&
  \bibinfo{author}{Aharonovich, I.}
\newblock \bibinfo{title}{Quantum emission from hexagonal boron nitride
  monolayers}.
\newblock \emph{\bibinfo{journal}{Nat. Nanotechnol.}}
  \textbf{\bibinfo{volume}{11}}, \bibinfo{pages}{37--41}
  (\bibinfo{year}{2016}).
\newblock \urlprefix\url{http://dx.doi.org/nnano.2015.242}.

\bibitem{10.1021/acsphotonics.8b00127}
\bibinfo{author}{Vogl, T.}, \bibinfo{author}{Campbell, G.},
  \bibinfo{author}{Buchler, B.~C.}, \bibinfo{author}{Lu, Y.} \&
  \bibinfo{author}{Lam, P.~K.}
\newblock \bibinfo{title}{Fabrication and deterministic transfer of
  high-quality quantum emitters in hexagonal boron nitride}.
\newblock \emph{\bibinfo{journal}{ACS Photonics}} \textbf{\bibinfo{volume}{5}},
  \bibinfo{pages}{2305--2312} (\bibinfo{year}{2018}).
\newblock \urlprefix\url{https://doi.org/10.1021/acsphotonics.8b00127}.

\bibitem{0022-3727-50-29-295101}
\bibinfo{author}{Vogl, T.}, \bibinfo{author}{Lu, Y.} \& \bibinfo{author}{Lam,
  P.~K.}
\newblock \bibinfo{title}{Room temperature single photon source using
  fiber-integrated hexagonal boron nitride}.
\newblock \emph{\bibinfo{journal}{J. Phys. D.}} \textbf{\bibinfo{volume}{50}},
  \bibinfo{pages}{295101} (\bibinfo{year}{2017}).
\newblock \urlprefix\url{http://stacks.iop.org/0022-3727/50/i=29/a=295101}.

\bibitem{doi:10.1021/acsnano.6b03602}
\bibinfo{author}{Tran, T.~T.} \emph{et~al.}
\newblock \bibinfo{title}{Robust multicolor single photon emission from point
  defects in hexagonal boron nitride}.
\newblock \emph{\bibinfo{journal}{ACS Nano}} \textbf{\bibinfo{volume}{10}},
  \bibinfo{pages}{7331--7338} (\bibinfo{year}{2016}).
\newblock \urlprefix\url{https://doi.org/10.1021/acsnano.6b03602}.

\bibitem{doi:10.1021/acsami.6b09875}
\bibinfo{author}{Choi, S.} \emph{et~al.}
\newblock \bibinfo{title}{Engineering and localization of quantum emitters in
  large hexagonal boron nitride layers}.
\newblock \emph{\bibinfo{journal}{ACS Appl. Mater. Interfaces}}
  \textbf{\bibinfo{volume}{8}}, \bibinfo{pages}{29642--29648}
  (\bibinfo{year}{2016}).
\newblock \urlprefix\url{https://doi.org/10.1021/acsami.6b09875}.

\bibitem{10.1038/nature23655}
\bibinfo{author}{Liao, S.-K.} \emph{et~al.}
\newblock \bibinfo{title}{Satellite-to-ground quantum key distribution}.
\newblock \emph{\bibinfo{journal}{Nature}} \textbf{\bibinfo{volume}{549}},
  \bibinfo{pages}{43 --47} (\bibinfo{year}{2017}).
\newblock \urlprefix\url{http://dx.doi.org/10.1038/nature23655}.

\bibitem{GSFC-STD-7000}
\bibinfo{author}{NASA}.
\newblock \bibinfo{title}{General environmental verification standard
  gsfc-std-7000} (\bibinfo{year}{2013}).
\newblock
  \urlprefix\url{https://standards.nasa.gov/standard/gsfc/gsfc-std-7000}.
\newblock
  \bibinfo{note}{Https://standards.nasa.gov/standard/gsfc/gsfc-std-7000}.

\bibitem{PhysRevB.93.205423}
\bibinfo{author}{Robert, C.} \emph{et~al.}
\newblock \bibinfo{title}{Exciton radiative lifetime in transition metal
  dichalcogenide monolayers}.
\newblock \emph{\bibinfo{journal}{Phys. Rev. B}} \textbf{\bibinfo{volume}{93}},
  \bibinfo{pages}{205423} (\bibinfo{year}{2016}).
\newblock \urlprefix\url{https://link.aps.org/doi/10.1103/PhysRevB.93.205423}.

\bibitem{doi:10.1021/acsphotonics.7b00086}
\bibinfo{author}{Kianinia, M.} \emph{et~al.}
\newblock \bibinfo{title}{Robust solid-state quantum system operating at 800
  k}.
\newblock \emph{\bibinfo{journal}{ACS Photonics}} \textbf{\bibinfo{volume}{4}},
  \bibinfo{pages}{768--773} (\bibinfo{year}{2017}).
\newblock \urlprefix\url{https://doi.org/10.1021/acsphotonics.7b00086}.

\bibitem{PhysRevB.98.081414}
\bibinfo{author}{Dietrich, A.} \emph{et~al.}
\newblock \bibinfo{title}{Observation of fourier transform limited lines in
  hexagonal boron nitride}.
\newblock \emph{\bibinfo{journal}{Phys. Rev. B}} \textbf{\bibinfo{volume}{98}},
  \bibinfo{pages}{081414} (\bibinfo{year}{2018}).
\newblock \urlprefix\url{https://link.aps.org/doi/10.1103/PhysRevB.98.081414}.

\bibitem{10.1088/1674-1056/20/8/086102}
\bibinfo{author}{Mai-Xing, H.} \emph{et~al.}
\newblock \bibinfo{title}{$\gamma$ radiation caused graphene defects and
  increased carrier density}.
\newblock \emph{\bibinfo{journal}{Chinese Phys. B}}
  \textbf{\bibinfo{volume}{20}}, \bibinfo{pages}{086102}
  (\bibinfo{year}{2011}).
\newblock \urlprefix\url{http://dx.doi.org/10.1088/1674-1056/20/8/086102}.

\bibitem{doi:10.1063/1.4963782}
\bibinfo{author}{Alexandrou, K.} \emph{et~al.}
\newblock \bibinfo{title}{Improving the radiation hardness of graphene field
  effect transistors}.
\newblock \emph{\bibinfo{journal}{Appl. Phys. Lett.}}
  \textbf{\bibinfo{volume}{109}}, \bibinfo{pages}{153108}
  (\bibinfo{year}{2016}).
\newblock \urlprefix\url{https://doi.org/10.1063/1.4963782}.

\bibitem{10.1002/pssa.201670681}
\bibinfo{author}{Walker, R.~C.}, \bibinfo{author}{Shi, T.},
  \bibinfo{author}{Silva, E.~C.}, \bibinfo{author}{Jovanovic, I.} \&
  \bibinfo{author}{Robinson, J.~A.}
\newblock \bibinfo{title}{Radiation effects on two-dimensional materials}.
\newblock \emph{\bibinfo{journal}{Phys. Status Solidi A}}
  \textbf{\bibinfo{volume}{213}}, \bibinfo{pages}{3268--3268}
  (\bibinfo{year}{2016}).
\newblock
  \urlprefix\url{https://onlinelibrary.wiley.com/doi/abs/10.1002/pssa.201670681}.

\bibitem{Milliron}
\bibinfo{author}{Milliron, R.}
\newblock \bibinfo{title}{Interorbital preps for neptune test launch - and
  eleven smallsats will go along for the ride...} (\bibinfo{year}{2017}).
\newblock
  \urlprefix\url{http://www.satmagazine.com/story.php?number=1600200139}.
\newblock \bibinfo{note}{Accessed 12/10/2018}.

\bibitem{10.1364/OPTICA.5.001128}
\bibinfo{author}{Proscia, N.~V.} \emph{et~al.}
\newblock \bibinfo{title}{Near-deterministic activation of room-temperature
  quantum emitters in hexagonal boron nitride}.
\newblock \emph{\bibinfo{journal}{Optica}} \textbf{\bibinfo{volume}{5}},
  \bibinfo{pages}{1128--1134} (\bibinfo{year}{2018}).
\newblock
  \urlprefix\url{http://www.osapublishing.org/optica/abstract.cfm?URI=optica-5-9-1128}.

\bibitem{spenvis}
\bibinfo{author}{{European Space Agency}}.
\newblock \bibinfo{title}{The space environment information system}.
\newblock \urlprefix\url{http://www.spenvis.oma.be/}.
\newblock \bibinfo{note}{Accessed 12/10/2018}.

\bibitem{bucik2000}
\bibinfo{author}{Bu{\v c}{\'i}k, R.} \emph{et~al.}
\newblock \bibinfo{title}{Distribution of gamma ray fluxes at altitude 500 km:
  Coronas-i data}.
\newblock \emph{\bibinfo{journal}{Acta Physica Slovaca}}
  \textbf{\bibinfo{volume}{50}}, \bibinfo{pages}{267} (\bibinfo{year}{2000}).

\bibitem{bucik1999}
\bibinfo{author}{Bu{\v c}{\'i}k, R.}, \bibinfo{author}{Dmitriev, A.},
  \bibinfo{author}{Kudela, K.} \& \bibinfo{author}{Ryumin, S.}
\newblock \bibinfo{title}{Gamma-radiation of the earth’s atmosphere from the
  coronas-i data}.
\newblock \emph{\bibinfo{journal}{Proc. 26 ICRC}} \textbf{\bibinfo{volume}{7}},
  \bibinfo{pages}{433--436} (\bibinfo{year}{1999}).

\bibitem{10.1038/nphoton2015.77}
\bibinfo{author}{Cassabois, G.}, \bibinfo{author}{Valvin, P.} \&
  \bibinfo{author}{B.Gil}.
\newblock \bibinfo{title}{Hexagonal boron nitride is an indirect bandgap
  semiconductor}.
\newblock \emph{\bibinfo{journal}{Nat. Photonics}}
  \textbf{\bibinfo{volume}{10}}, \bibinfo{pages}{262--266}
  (\bibinfo{year}{2016}).
\newblock \urlprefix\url{http://dx.doi.org/10.1038/nphoton2015.77}.

\bibitem{10.1038/lsa.2016.46}
\bibinfo{author}{Yang, J.} \emph{et~al.}
\newblock \bibinfo{title}{Atomically thin optical lenses and gratings}.
\newblock \emph{\bibinfo{journal}{Light Sci. Appl.}}
  \textbf{\bibinfo{volume}{5}}, \bibinfo{pages}{e16046} (\bibinfo{year}{2002}).
\newblock \urlprefix\url{http://dx.doi.org/10.1038/lsa.2016.46}.

\bibitem{doi:10.1021/nl3026357}
\bibinfo{author}{Gutiérrez, H.~R.} \emph{et~al.}
\newblock \bibinfo{title}{Extraordinary room-temperature photoluminescence in
  triangular ws$_2$ monolayers}.
\newblock \emph{\bibinfo{journal}{Nano Lett.}} \textbf{\bibinfo{volume}{13}},
  \bibinfo{pages}{3447--3454} (\bibinfo{year}{2013}).
\newblock \urlprefix\url{https://doi.org/10.1021/nl3026357}.

\bibitem{toi}
\bibinfo{author}{Chu, S. Y.~F.}, \bibinfo{author}{Ekstr\"om, L.~P.} \&
  \bibinfo{author}{Firestone, R.~B.}
\newblock \bibinfo{title}{{WWW Table of Radioactive Isotopes}}.
\newblock \urlprefix\url{http://nucleardata.nuclear.lu.se/nucleardata/toi/}.
\newblock \bibinfo{note}{Database version 28/02/1999}.

\bibitem{10.1134/S2075113316020040}
\bibinfo{author}{Anikeyev, V.~V.} \emph{et~al.}
\newblock \bibinfo{title}{Effect of electron irradiation on the formation and
  healing of defects in carbon nanotubes}.
\newblock \emph{\bibinfo{journal}{Inorganic Mater: Appl. Research}}
  \textbf{\bibinfo{volume}{7}}, \bibinfo{pages}{204--209}
  (\bibinfo{year}{2016}).
\newblock \urlprefix\url{https://doi.org/10.1134/S2075113316020040}.

\bibitem{10.1038/ncomms7293}
\bibinfo{author}{Hong, J.} \emph{et~al.}
\newblock \bibinfo{title}{Exploring atomic defects in molybdenum disulphide
  monolayers}.
\newblock \emph{\bibinfo{journal}{Nat. Commun.}} \textbf{\bibinfo{volume}{6}},
  \bibinfo{pages}{6293} (\bibinfo{year}{2015}).
\newblock \urlprefix\url{http://dx.doi.org/10.1038/ncomms7293}.

\bibitem{doi:10.1002/anie.201508828}
\bibinfo{author}{Liu, Y.}, \bibinfo{author}{Stradins, P.} \&
  \bibinfo{author}{Wei, S.-H.}
\newblock \bibinfo{title}{Air passivation of chalcogen vacancies in
  two-dimensional semiconductors}.
\newblock \emph{\bibinfo{journal}{Angew. Chem. Int. Ed.}}
  \textbf{\bibinfo{volume}{55}}, \bibinfo{pages}{965--968}.
\newblock
  \urlprefix\url{https://onlinelibrary.wiley.com/doi/abs/10.1002/anie.201508828}.

\bibitem{doi:10.1021/acs.nanolett.5b00952}
\bibinfo{author}{Lu, J.} \emph{et~al.}
\newblock \bibinfo{title}{Atomic healing of defects in transition metal
  dichalcogenides}.
\newblock \emph{\bibinfo{journal}{Nano Lett.}} \textbf{\bibinfo{volume}{15}},
  \bibinfo{pages}{3524--3532} (\bibinfo{year}{2015}).
\newblock \urlprefix\url{https://doi.org/10.1021/acs.nanolett.5b00952}.

\bibitem{10.1364/AO.55.006251}
\bibinfo{author}{Wei, K.}, \bibinfo{author}{Liu, Y.}, \bibinfo{author}{Yang,
  H.}, \bibinfo{author}{Cheng, X.} \& \bibinfo{author}{Jiang, T.}
\newblock \bibinfo{title}{Large range modification of exciton species in
  monolayer ws2}.
\newblock \emph{\bibinfo{journal}{Appl. Opt.}} \textbf{\bibinfo{volume}{55}},
  \bibinfo{pages}{6251--6255} (\bibinfo{year}{2016}).
\newblock \urlprefix\url{http://ao.osa.org/abstract.cfm?URI=ao-55-23-6251}.

\bibitem{10.1038/nphoton.2013.179}
\bibinfo{author}{Miyauchi, Y.} \emph{et~al.}
\newblock \bibinfo{title}{Brightening of excitons in carbon nanotubes on
  dimensionality modification}.
\newblock \emph{\bibinfo{journal}{Nat. Photon.}} \textbf{\bibinfo{volume}{7}},
  \bibinfo{pages}{715--719} (\bibinfo{year}{2013}).
\newblock \urlprefix\url{http://dx.doi.org/10.1038/nphoton.2013.179}.

\bibitem{PhysRevB.54.11169}
\bibinfo{author}{Kresse, G.} \& \bibinfo{author}{Furthm\"uller, J.}
\newblock \bibinfo{title}{Efficient iterative schemes for ab initio
  total-energy calculations using a plane-wave basis set}.
\newblock \emph{\bibinfo{journal}{Phys. Rev. B}} \textbf{\bibinfo{volume}{54}},
  \bibinfo{pages}{11169--11186} (\bibinfo{year}{1996}).
\newblock \urlprefix\url{https://link.aps.org/doi/10.1103/PhysRevB.54.11169}.

\bibitem{PhysRevB.59.1758}
\bibinfo{author}{Kresse, G.} \& \bibinfo{author}{Joubert, D.}
\newblock \bibinfo{title}{From ultrasoft pseudopotentials to the projector
  augmented-wave method}.
\newblock \emph{\bibinfo{journal}{Phys. Rev. B}} \textbf{\bibinfo{volume}{59}},
  \bibinfo{pages}{1758--1775} (\bibinfo{year}{1999}).
\newblock \urlprefix\url{https://link.aps.org/doi/10.1103/PhysRevB.59.1758}.

\bibitem{PhysRevLett.78.1396}
\bibinfo{author}{Perdew, J.~P.}, \bibinfo{author}{Burke, K.} \&
  \bibinfo{author}{Ernzerhof, M.}
\newblock \bibinfo{title}{Generalized gradient approximation made simple [phys.
  rev. lett. 77, 3865 (1996)]}.
\newblock \emph{\bibinfo{journal}{Phys. Rev. Lett.}}
  \textbf{\bibinfo{volume}{78}}, \bibinfo{pages}{1396--1396}
  (\bibinfo{year}{1997}).
\newblock \urlprefix\url{https://link.aps.org/doi/10.1103/PhysRevLett.78.1396}.

\bibitem{10.1016/j.carbon.2010.12.057}
\bibinfo{author}{Mathew, S.} \emph{et~al.}
\newblock \bibinfo{title}{The effect of layer number and substrate on the
  stability of graphene under mev proton beam irradiation}.
\newblock \emph{\bibinfo{journal}{Carbon}} \textbf{\bibinfo{volume}{49}},
  \bibinfo{pages}{1720 -- 1726} (\bibinfo{year}{2011}).
\newblock
  \urlprefix\url{http://www.sciencedirect.com/science/article/pii/S000862231000936X}.

\bibitem{PhysRevLett.109.035503}
\bibinfo{author}{Komsa, H.-P.} \emph{et~al.}
\newblock \bibinfo{title}{Two-dimensional transition metal dichalcogenides
  under electron irradiation: Defect production and doping}.
\newblock \emph{\bibinfo{journal}{Phys. Rev. Lett.}}
  \textbf{\bibinfo{volume}{109}}, \bibinfo{pages}{035503}
  (\bibinfo{year}{2012}).
\newblock
  \urlprefix\url{https://link.aps.org/doi/10.1103/PhysRevLett.109.035503}.

\bibitem{doi:10.1021/acsami.8b07506}
\bibinfo{author}{Ngoc My~Duong, H.} \emph{et~al.}
\newblock \bibinfo{title}{Effects of high-energy electron irradiation on
  quantum emitters in hexagonal boron nitride}.
\newblock \emph{\bibinfo{journal}{ACS Appl. Mater. Interfaces}}
  \textbf{\bibinfo{volume}{10}}, \bibinfo{pages}{24886--24891}
  (\bibinfo{year}{2018}).
\newblock \urlprefix\url{https://doi.org/10.1021/acsami.8b07506}.

\bibitem{10.1016/j.nimb.2010.02.091}
\bibinfo{author}{Ziegler, J.~F.}, \bibinfo{author}{Ziegler, M.} \&
  \bibinfo{author}{Biersack, J.}
\newblock \bibinfo{title}{Srim - the stopping and range of ions in matter}.
\newblock \emph{\bibinfo{journal}{Nucl. Instr. Meth. Phys. Res. B}}
  \textbf{\bibinfo{volume}{268}}, \bibinfo{pages}{1818 -- 1823}
  (\bibinfo{year}{2010}).
\newblock
  \urlprefix\url{http://www.sciencedirect.com/science/article/pii/S0168583X10001862}.

\bibitem{ESTAR}
\bibinfo{author}{Berger, M.}, \bibinfo{author}{Coursey, J.},
  \bibinfo{author}{Zucker, M.} \& \bibinfo{author}{Chang, J.}
\newblock \bibinfo{title}{{ESTAR, PSTAR, and ASTAR}: Computer programs for
  calculating stopping-power and range tables for electrons, protons, and
  helium ions, (version 2.0.1)} (\bibinfo{year}{2005}).
\newblock \urlprefix\url{http://physics.nist.gov/Star}.
\newblock \bibinfo{note}{Http://physics.nist.gov/Star, National Institute of
  Standards and Technology, Gaithersburg, MD}.

\bibitem{doi:10.1002/sca.20000}
\bibinfo{author}{Drouin, D.} \emph{et~al.}
\newblock \bibinfo{title}{Casino v2.42 - a fast and easy-to-use modeling tool
  for scanning electron microscopy and microanalysis users}.
\newblock \emph{\bibinfo{journal}{Scanning}} \textbf{\bibinfo{volume}{29}},
  \bibinfo{pages}{92--101} (\bibinfo{year}{2007}).
\newblock
  \urlprefix\url{https://onlinelibrary.wiley.com/doi/abs/10.1002/sca.20000}.

\bibitem{PhysRevB.85.085201}
\bibinfo{author}{Popescu, V.} \& \bibinfo{author}{Zunger, A.}
\newblock \bibinfo{title}{Extracting e versus k effective band structure from
  supercell calculations on alloys and impurities}.
\newblock \emph{\bibinfo{journal}{Phys. Rev. B}} \textbf{\bibinfo{volume}{85}},
  \bibinfo{pages}{085201} (\bibinfo{year}{2012}).
\newblock \urlprefix\url{https://link.aps.org/doi/10.1103/PhysRevB.85.085201}.

\bibitem{PyVaspwfc}
\bibinfo{author}{Zheng, Q.}
\newblock \bibinfo{title}{{VASP Band Unfolding}}.
\newblock
  \bibinfo{howpublished}{\url{https://github.com/QijingZheng/VaspBandUnfolding}}
  (\bibinfo{year}{2018}).

\bibitem{PhysRevLett.51.1884}
\bibinfo{author}{Perdew, J.~P.} \& \bibinfo{author}{Levy, M.}
\newblock \bibinfo{title}{Physical content of the exact kohn-sham orbital
  energies: Band gaps and derivative discontinuities}.
\newblock \emph{\bibinfo{journal}{Phys. Rev. Lett.}}
  \textbf{\bibinfo{volume}{51}}, \bibinfo{pages}{1884--1887}
  (\bibinfo{year}{1983}).
\newblock \urlprefix\url{https://link.aps.org/doi/10.1103/PhysRevLett.51.1884}.

\bibitem{10.1088/2053-1583/4/1/015026}
\bibinfo{author}{Zhang, C.} \emph{et~al.}
\newblock \bibinfo{title}{Systematic study of electronic structure and band
  alignment of monolayer transition metal dichalcogenides in van der waals
  heterostructures}.
\newblock \emph{\bibinfo{journal}{2D Materials}} \textbf{\bibinfo{volume}{4}},
  \bibinfo{pages}{015026} (\bibinfo{year}{2017}).
\newblock \urlprefix\url{http://stacks.iop.org/2053-1583/4/i=1/a=015026}.

\end{thebibliography}
\end{document}